\documentclass{aastex}
\usepackage{spr-astr-addons}
\usepackage{color}

\def\nh{$n_{\mathrm{H}}$\/}
\def\lledd{$L/L_{\rm Edd}$}

\def\rfe{$R_{\rm FeII}$}

\def\msol{M$_\odot$\/}

\def\rg{R$_{\rm g}$\/}
\def\ltsima{$\; \buildrel < \over \sim \;$}
\def\ltsim{\lower.5ex\hbox{\ltsima}}  
\def\simlt{\lower.5ex\hbox{\ltsima}}  
\def\gtsima{$\; \buildrel > \over \sim \;$}

\def\gtsim{\lower.5ex\hbox{\gtsima}} 
\def\simgt{\lower.5ex\hbox{\gtsima}}

\def\ha{{\sc H}$\alpha$}
\def\lya{{ Ly}$\alpha$}
\def\civ{{\sc{Civ}}$\lambda$1549\/}

\def\cmq{cm$^{-2}$\/}
\def\cm3{cm$^{-3}$\/}
\def\hb{{\sc{H}}$\beta$\/}

\def\hbbc{{\sc{H}}$\beta_{\rm BC}$\/}

\def\hbnc{{\sc{H}}$\beta_{\rm NC}$\/}
\def\mgii{{Mg\sc{ii}}$\lambda$2800\/}

\def\o4363{{\sc{[Oiii]}}$\lambda$4363\/}
\def\caii{{Ca{\sc ii}}}

\def\feiiopt{{Fe \sc{ii}}$_{\rm opt}$\/}
\def\feii{{Fe\sc{ii}}\/}

\def\fe{{\sc{Fe}}\/}
\def\gs{$\Gamma_\mathrm{S}$\/}

\def\fe76087{{\sc [Fe vii]}$\lambda$6087\/}
\def\oiii{{\sc [Oiii]}$\lambda$5007}

\def\kms{km~s$^{-1}$}
\def\rk{$R_{\rm K}$\/}
\def\ergss{ergs s$^{-1}$\/}

\def\rk{{$R{\rm _K}$}\/}

\RequirePackage{color}

\begin{document}

\title{Balmer line shifts in quasars}
\shorttitle{Short article title}
\shortauthors{Autors et al.}

\author{J. W. Sulentic}
\affil{Instituto de Astrof{\'\i}sica de Andaluc{\'\i}a (CSIC),    Granada, Spain}
\and
\author{P. Marziani}
\affil{INAF, Osservatorio Astronomico di Padova, Padova, Italia}
\and
\author{A. Del Olmo}
\affil{Instituto de Astrof{\'\i}sica de Andaluc{\'\i}a (CSIC),    Granada, Spain}
\and 
\author{S. Zamfir}
\affil{University of Wisconsin,  Stevens Point, WI, USA}



\defcitealias{zamanovetal02}{Z02}

\begin{abstract}
We offer a broad review of Balmer line phenomenology in type 1 active galactic nuclei, briefly summarising luminosity and radio loudness effects, and discussing interpretation in terms of nebular physics along the 4D eigenvector 1 sequence of quasars. We stress that relatively rare, peculiar Balmer line profiles  (i.e., with large shifts with respect to the rest frame or double and multiple peaked) that start attracted attentions since the 1970s are still passable of multiple dynamical interpretation. More mainstream objects are  still not fully understood as well, since competing dynamical models and geometries are possible. Further progress may come from inter-line comparison across the 4D Eigenvector 1 sequence.
\end{abstract}

\keywords{galaxies: active; quasars: emission lines; quasars: general}

\section{Introduction: Quasars and line shifts}
\label{intro}

Internal line shifts in quasar spectra have played an ever-increasing role in our understanding of quasar 
structure and dynamics. Apart from early attempts to explain quasar cosmological redshifts as intrinsic to 
the quasar \citep[see e.g., the historical account in Chapter 2 of ][]{donofrioetal12}  that lead to 
serious inconsistency  \citep{greensteinschmidt64}, the realization that all lines in a quasar do not yield 
the same  redshift was slow to come. Until the early 1980s and the pioneering paper by Martin Gaskell 
\citep{gaskell82}    {\em redshift difference between emission lines had not been systematically studied 
in quasars, } even if some  cases of broad line shifts ($\sim 1000$ \kms with respect to narrow lines) 
had been noted in Seyfert galaxies \citep{osterbrock79}.  Gaskell's  results on the difference between 
low- and high-ionization lines (LILs and HILs, represented by \mgii\ and \civ\ respectively) have remained 
valid since then, and have yielded the observational basis for a model that is still today providing a basic 
interpretative sketch for the structure of the broad line emitting region \citep{collinsouffrinetal88}. In 
that model, HIL blue shifts are associated with radial motion in a flattened structure (the accretion disk) 
emitting mainly LILs.  Later developments  \citep{tytlerfan92,marzianietal96,corbinboroson96} confirmed the 
earlier results.  \citet{sulenticetal95} suggested that results were not the same for all quasars, and that a 
first distinction between radio-loud and radio-quiet quasars was necessary: blueshifts were mostly present  
in radio quiet (RQ) quasars, but definitely rarer or of lower amplitudes in radio loud (RL). This finding (confirmed in \citealt{marzianietal96} 
and \citealt{richardsetal11}) has been further contextualised within the Pop. A and B concept in the
4D Eigenvector 1 formalism \citep[e.g.][]{sulenticetal00a,sulenticetal07,sulenticetal11}. In the following we 
will focus attention on the \hb\ line, as representative of LILs, and its contextualization within 4DE1 (\S \ref{4de1}).
We   report the observational properties in a systematic way (\S \ref{phen}) that includes a robust empirical 
interpretation based on nebular physics (\S \ref{ei}), and effects of luminosity and radio loudness (\S \ref{effects}). 
We then discuss the physical processes that may be relevant in line profile broadening (\S \ref{physics}). 
We conclude with remarks on unknown aspects and desiderata for further observations (\S \ref{build} and \S\ \ref{open}).

\section{(4D)Eigenvector 1 quasar contextualization at low-$z$}
\label{4de1}

The (4D) Eigenvector 1 quasar con\-text\-ualiz\-ation sch\-eme at low-$z$ offers a powerful tool for
interpretation of the \hb\  emission line (and of LILs in general)  profiles.  Eigenvector 1 was
originally defined from a Principal Component Analysis (PCA) of 87 PG quasars involving an anticorrelation 
between optical FeII intensity, half-maximum profile width of \hb\  and peak intensity of \oiii\ \citep{borosongreen92}.  
E1 was afterwards expanded to 4DE1 with the addition of X-ray photon index and \civ\ profile
shift measures \citep{sulenticetal00a,sulenticetal08,sulenticetal11}. 4DE1 allows one to define a quasar main 
sequence in 4DE1 space \citep{sulenticetal00a,marzianietal01}.  

The principle parameters of the 4DE1 space are:
 
 \begin{itemize}
 \item Full width at half maximum of broad \hb\ (FWHM \hb) which is our diagnostic low ionization lines 
observable from the ground out to z $\approx$ 0.9 with optical and (intermittently to $z = 3.7$) with 
infrared spectroscopy. It (and/or FWHM \mgii) is thought to be a measure of assumed virialized motions 
in the AD and is thus crucial for estimating black hole masses for large samples of quasars.
\item  Ratio of the equivalent width of the 4570 \AA\ optical FeII blend and broad \hb\ (\rfe = W 4570 
FeII blend/W \hb ). It is sensitive to the ionization state, the electron density and column density of 
the BLR gas arising, as far as we can tell, in (at least part of) the gas that produces 
\hb\ \citep[e.g.,][]{huetal08,marzianietal10}. \item Centroid shift at FWHM of the high ionization 
line \civ\ (at half, maximum, c(1/2)). It is a strong diagnostic of winds/outflows 
\citep[e.g.,][]{marzianietal96,leighlymoore04}. \item  Soft X-ray photon index (\gs). Most likely an indicator 
of optically thick Comptonized radiation, and probably a diagnostic of the thermal emission from the AD 
or of a Compton-thick corona above the disk \citep[][]{doneetal12,haardtmaraschi91,jinetal09}.
\end{itemize}

The 4DE1 approach allows the definition of spectral types following source occupation in the optical
4DE1 plane FWHM(H$\beta$) vs \rfe\ \citep{sulenticetal02} and can be extended to include all four dimensions  
\citep[e.g., ][for \civ]{bachevetal04}. It is possible to identify two populations along the sequence: Population
A with FWHM(H$\beta$)$\le 4000$ \kms, and a second Population B of sources with broader lines \citep[see ][for the 
rationale behind two distinct quasar populations]{sulenticetal07,sulenticetal11}. 

\section{The phenomenology of the \hb\ emission line profile in the 4DE1 context}
\label{phen}

It is customary to distinguish between broad and narrow components of Balmer lines, with Type 1 AGN showing 
a non-zero broad component \citep[the case closest to type 2 is type 1.9, for which the BC is visible in 
\ha\ but not in \hb][]{osterbrock81}. It is widely believed that the emitting region of these sources suffers 
heavy obscuration \citep{antonuccimiller85,antonucci93}. In some cases -- but apparently not in all sources 
\citep{pappaetal01,wolteretal05,bianchietal08,petrov09} -- a BC becomes visible in polarised light.  
In the following we restrict attention to sources that do not show strong evidence of obscuration in their 
line profiles (i.e., we will exclude type 1.5, 1.8 and 1.9 sources of the old Osterbrock's classification).

\begin{figure*}[htp!]
\includegraphics[scale=0.1]{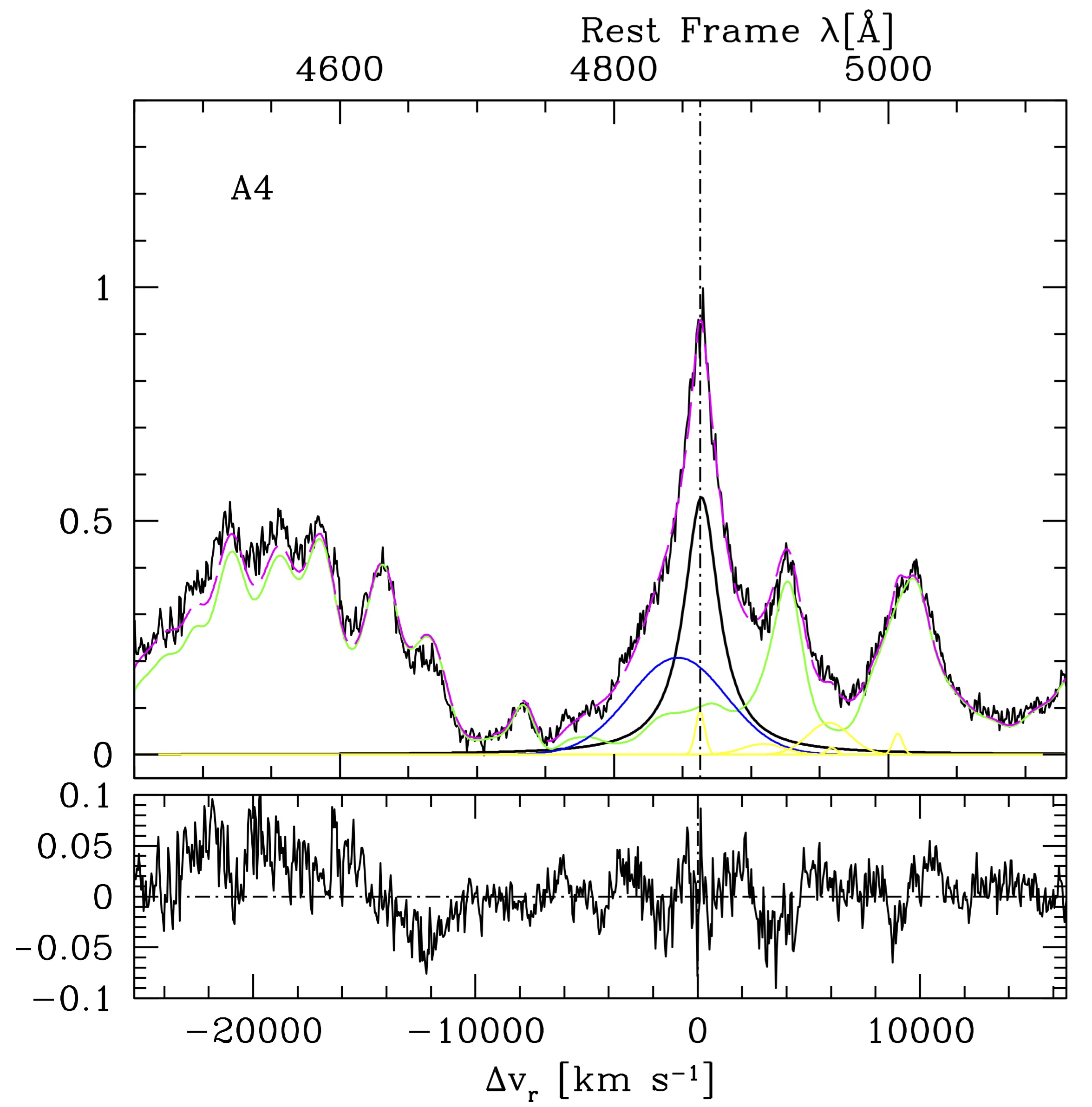}
\includegraphics[scale=0.1]{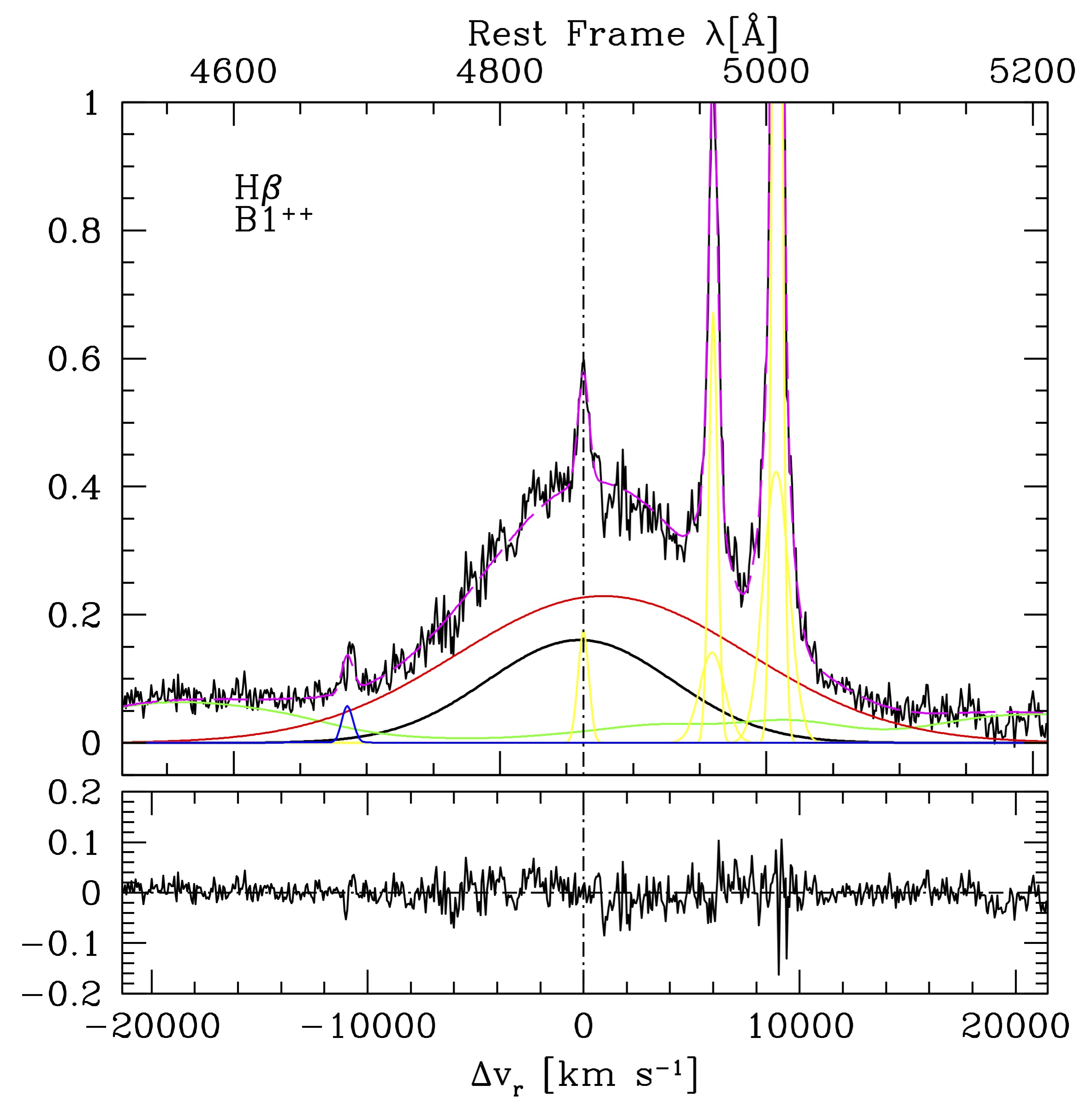}
\caption{Median composite \hb\ profiles at the extreme ends of the 4DE1 sequence: spectral type A4 (left), spectral type (B1++), from the sample of \citet{marzianietal13a}. The full specfit model of the spectrum is traced by the magenta dotted line. Narrow lines are colored yellow, the BC black, the blue shifted component blue, and the VBC red. The \feiiopt\ emission (very strong in A4 but almost undetectable in B1++) is coloured pale green. Abscissae are rest frame wavelength in \AA\ and radial velocity from the peak position of \hbnc. The bottom panels show the difference between the spectrum and its model. }
\label{fig:e1}        
\end{figure*}

Before 4DE1, several workers provided excellent data on \hb\  \citep[e.g.,][]{shuder81,stirpe90} but their 
interpretation was limited by the lack of  any contextualization. Without contextualization every quasar is an 
island under itself without a clear understanding of how it fits into the general population. The FWHM \hb\ distribution 
is puzzling since its includes profiles over a very wide range of widths, from less than 1000 to more than 
20000 \kms. These line width limits can be related to physical processes, although under several assumptions  
\citep{laor03} that may be  justified  only after a first organization of empirical properties is accomplished. A 
first attempt to describe in a systematic way the shape of broad lines in low-$z$\ quasars devised a qualitative 
assignment based on peak shift and shape asymmetry:  for example, a redward asymmetric profile with a redshifted 
peak would be classified as AR,R, a symmetric unshifted as S, etc. \citep{sulentic89}. This and more quantitative 
analyses of the line wing shape stressed  the systematic changes occurring in large samples 
\citep{borosongreen92,robinson95}.  Further  work of our group involved the analysis of observations collected 
over a decade and presented in \citet{marzianietal03a}. Results from the analysis of those data were later 
confirmed by a larger sample of 470 bright quasars from SDSS DR5 \citep{zamfiretal10}. The main findings can be 
summarised as follows:

\begin{itemize}
\item  There is  a trend in \hb\ line profile properties that is correlated with the 4DE1 sequence \citep{sulenticetal07}. 
Line profile shapes change along the sequence, from a Lorentzian with a blue shifted excess at the line base to a redward 
asymmetric profile best fit with 2 Gaussian components. The latter profile is typical of the larger 
FWHM $\gtrsim 4000$ \kms. Fig. \ref{fig:e1} shows the spectral types at the opposite ends of the sequence. In spectral 
type A4 (at the high FeII end of the 4ED1 sequence)  the blue excess of \hb\ is especially evident, as the shallow, redward very broad profile is for B1++ (at the opposite end of the sequence, with broadest lines and very weak FeII emission).
\item Peculiar, double- or multiple peaked profiles are more frequent toward the extreme Pop. B spectral types  
i.e., B1++ with FWHM (\hb) $\ge$ 12000 \kms.  
\item Radio loud sources are predominantly Pop. B \citep{marzianietal96,marzianietal01,sulenticetal03,rokakietal03,zamfiretal08}; 
radio core-dominated (CD) sources show systematically narrower profiles (and distribute in 4DE1 spaced like radio-quiet sources), 
while lobe-dominated (LD assumed to be Fanaroff-Riley II,FRII) are frequently  associated with very broad profiles (almost exclusively Population B). 
\end{itemize}

\begin{figure}[htp!]
\includegraphics[scale=0.09]{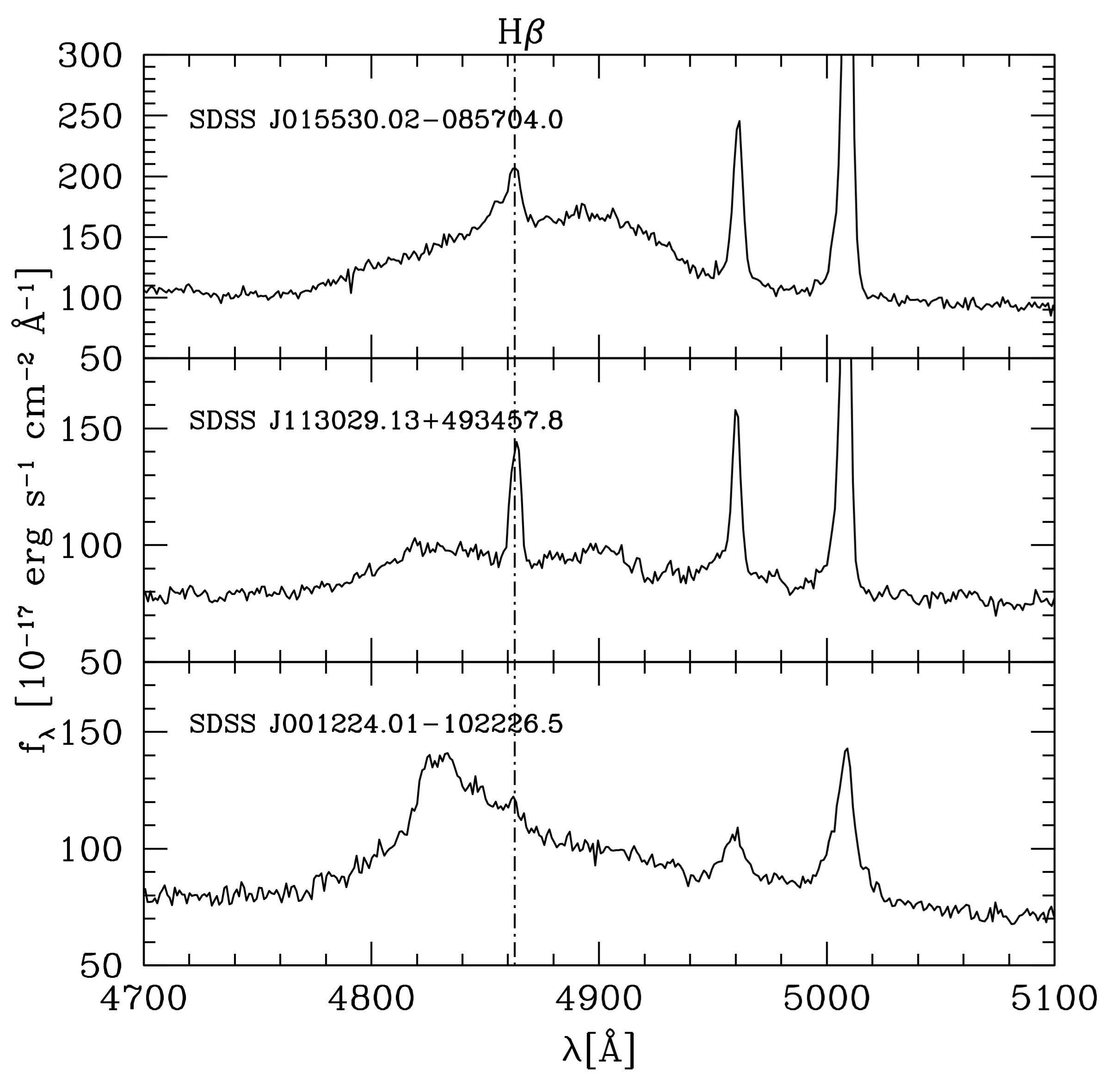}
\caption{Three examples of peculiar \hb\ profiles from the sample of \citet{zamfiretal10}.  Abscissa is rest frame wavelength in \AA, ordinate is specific flux in 10$^{-17}$ \ergss\ \cmq\ \AA$^{-1}$.   }
\label{fig:pec}        
\end{figure}

Peculiar profiles can be empirically grouped in three large categories: (1) double peaked profiles with small peak separation 
($\ltsim 10^{3}$ \kms).  These sources are apparently rare, since only a  handful are known. We stress the apparently since some cases could be masked by the blending of \hbbc\ and  \hbnc. Prototypes include Akn 120, IC 4329A, 
and OX 169 \citep{marzianietal92}; (2) wide separation double peaked profiles ($\Delta v \gtsim 10^{3}$ \kms), whose prototype 
are Arp 120B and 3C 390.3 \citep{chenetal89,zhang11}. An example is shown in Fig. \ref{fig:pec}. (3) Single peaked profiles 
with a large shift, either to the blue or to the red (two examples, one for each case, are also shown in Fig. \ref{fig:pec}). A prototypical source, Mkn 668 \citep{marzianietal93}, shows a peak displacement of $+3000$ \kms. \citet{stratevaetal03} found 
that class (2) profiles are also relatively rare ($\simlt 2$\%, considering that most of the \citealt{stratevaetal03} sources are not actually double-peaked sources). The prevalence is higher among RL   but significant also among RQ sources.

\subsection{Empirical Interpretation of the \hb\ emission line profile along the 4DE1 sequence}
\label{ei}

Attempts at interpretation of single emission lines are doomed to provide rather ambiguous results. The emitting region is not spatially resolved, and the profiles are usually smooth. They can be described by simple functional forms (logarithmic, or 
Lorentzian), and these profile can be  modelled assuming widely different structural and dynamical conditions. On the converse, comparison of representative HILs to  representative LILs (\hb\ and \mgii) provides important insights \citep{marzianietal96}. Some \hb\ profiles show a significant excess with respect to a symmetric Lorentzian especially in the spectral types associated with the highest \lledd\ \citep{marzianietal13}. The blue shifted emission has been interpreted as an outflowing component which has a physical correspondence 
if there is e.g. a non-rotating radiation driven wind \citep[][]{elvis00}.  

\begin{figure}[htp!]
\includegraphics[scale=0.15]{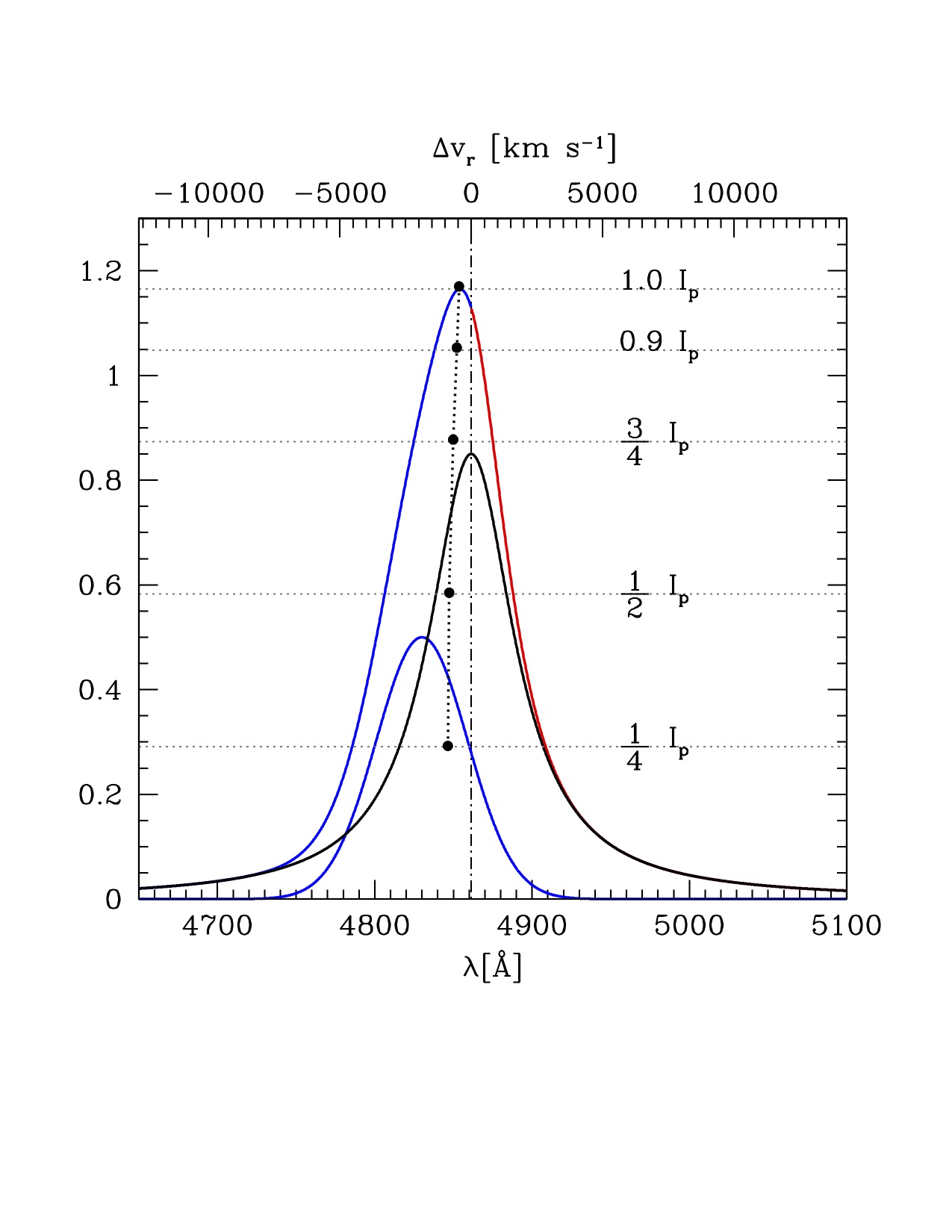}
\includegraphics[scale=0.15]{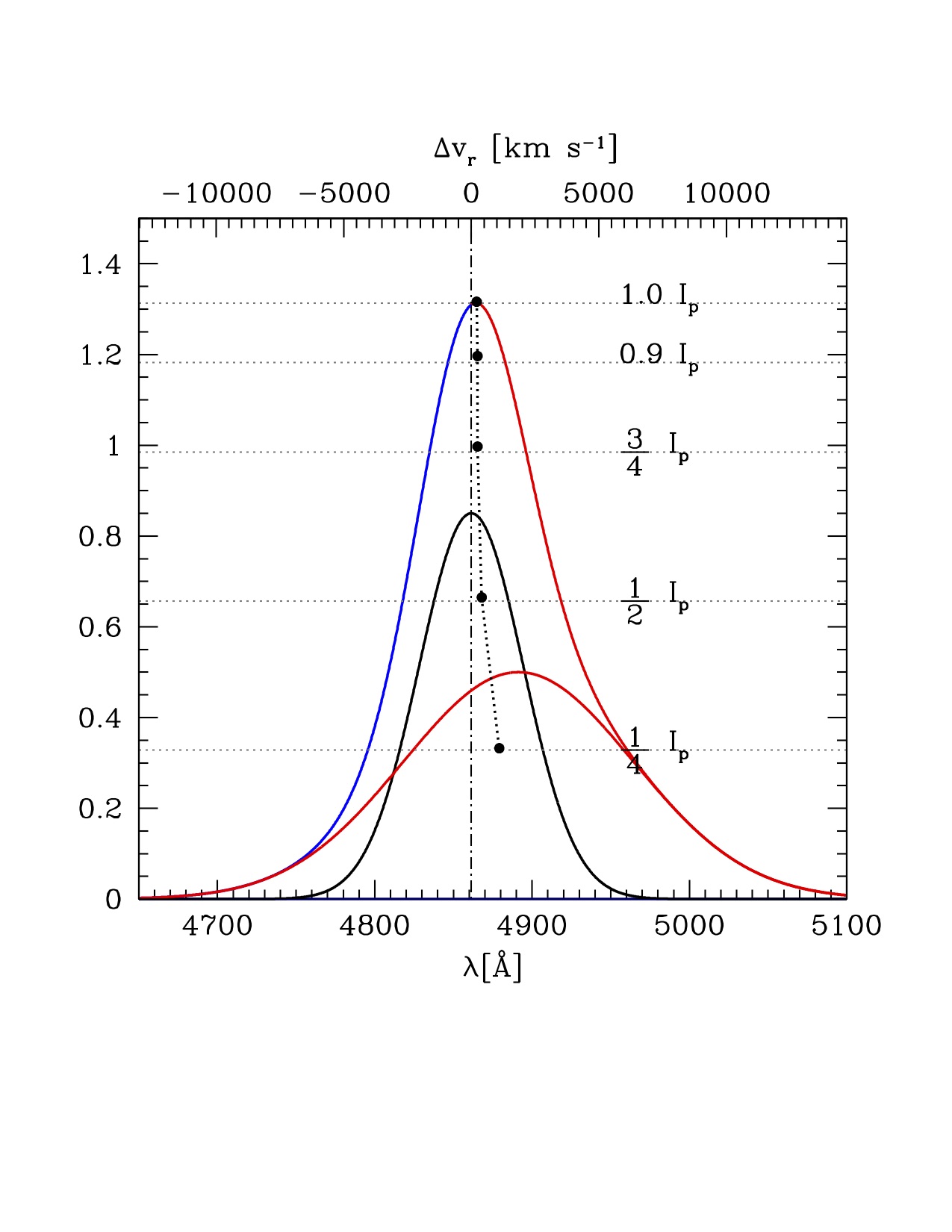}
\caption{Mock \hb\ profile illustrating the line decomposition into a BC and a blueshifted  component for Pop. A (top), and into a BC, VBC and blue shifted component for Pop. B (bottom). Radial velocities are measured for the peak intensity $I_\mathrm{p}$,  and the centroids (black spots) at different fractional intensities on the {\em full} profile, to obtain a quantitative parameterisation that is independent of the decomposition of the profile.   }
\label{fig:mock}        
\end{figure}

Further  interpretation  of the broad profiles involves ``stratification'' of the emitting region: we associate (likely an over-simplification of a continuous radial trend) the broad component  with a lower ionization Broad Line Region (BLR), where line broadening is predominantly virial, and where \feii, \caii\ are also emitted \citep[][and references therein]{martinez-aldamaetal15}.  Consistently, there is no significant shift ($\ltsim 200$ \kms) between \hb\ BC and \feii\ optical emission 
\citep{sulenticetal12}. The very broad component is associated with an inner region of higher 
ionization, the Very Broad Line Region (VBLR) emitting no \feii\ and showing lower continuum responsivity \citep{sneddengaskell07,goadkorista14}. The typical decomposition BC/VBC is shown by the mock profile in Fig. \ref{fig:mock} as well as by the composite profile in Fig. \ref{fig:e1}. LIL profiles often show asymmetries but the amplitude of the asymmetries and shifts is usually much less than their FWHM \citep{sulentic89,marzianietal03a,zamfiretal10}.  The stratification view is consistent with  virial motions dominating both the core and the wings of the Pop. B LILs, although the redward asymmetry of 
Balmer and other LILs is difficult to explain.  

\subsection{Luminosity and radio-loudness effects on the \hb\ line profile}
\label{effects}

\begin{figure}[htp!]
\includegraphics[scale=0.09]{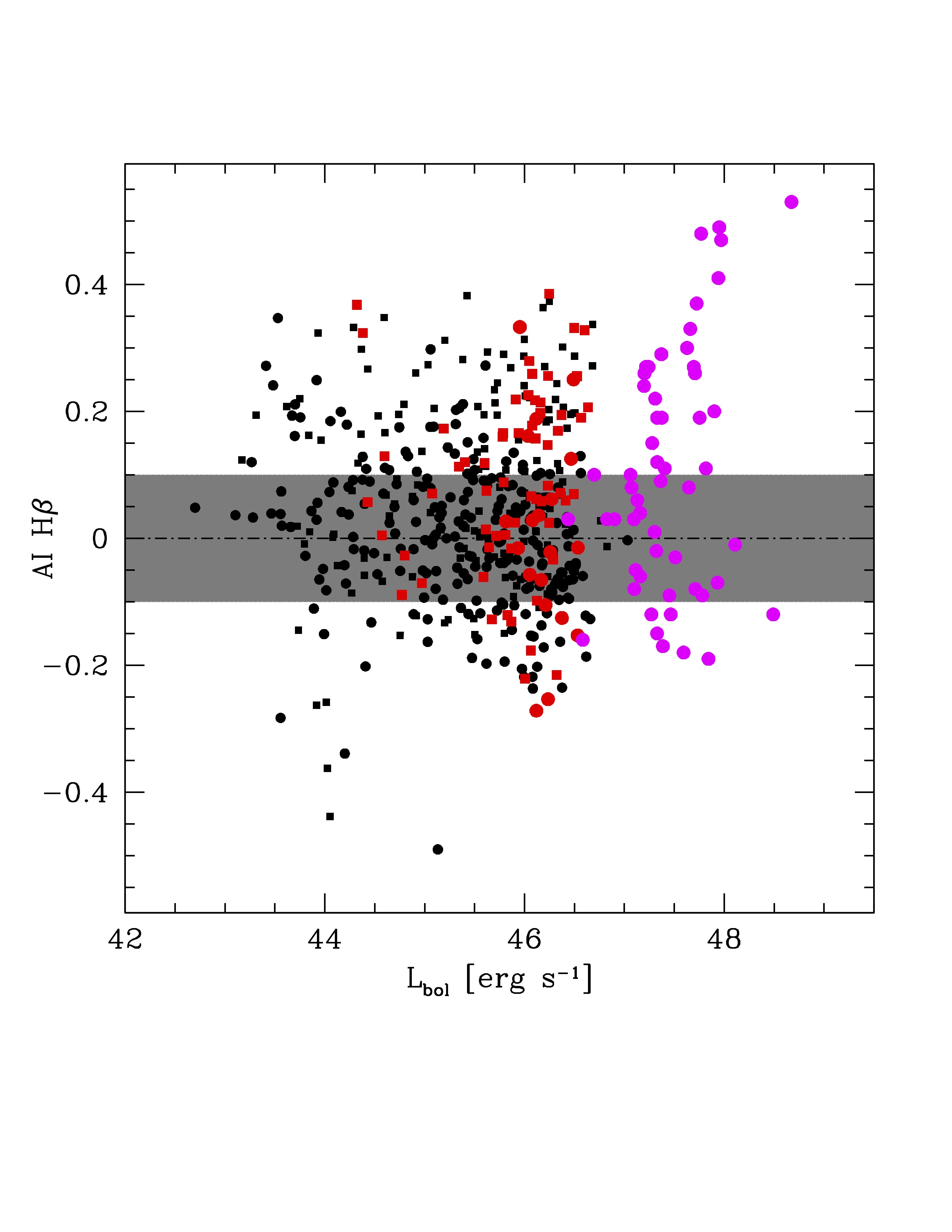}
\caption{Luminosity effects on line profiles shapes: the \hb\ AI behaviour  as a function of bolometric luminosity in \ergss. Black symbols: RQ, red symbols: RL, squares: Pop. B, circles: Pop. A. from the sample of \citet{zamfiretal10}. The magenta points at high $L_\mathrm{bol}$\ represent the sample of \citet{marzianietal09}. The shaded band width is set  by the uncertainty at a $\pm2 \sigma$\ confidence level in individual AI measurements. Within this band, profiles can be considered symmetric.  }
\label{fig:lum}        
\end{figure}

The main luminosity effect on \hb\ is a systematic increase in the minimum FWHM \citep{marzianietal09} 
which has been shown in several papers since 2009 and will not be shown again. At the same time we 
see that shifts toward the red close to the line base, at 1/4 fractional intensity become more 
frequent in higher $L$\ sources (this is reflected in an increased redward asymmetry). At 
present we do not see any strong RL effect on \hb. The median \hb\ line profiles for RL and 
RQ Pop. B sources are practically indistinguishable \citep{marzianietal03b,zamfiretal08}. Extreme 
cases of  redward asymmetry are found among radio loud quasars but there is no doubt that  
Pop. B RQ sources are also capable of impressive redward asymmetries (especially at high luminosity) 
with AI$\approx$ 0.5. Most of the quasars in our high $L$\  sample \citep{marzianietal09} 
are RQ quasars (Fig. \ref{fig:lum}). 

\begin{figure}[htp!]
\includegraphics[scale=0.09]{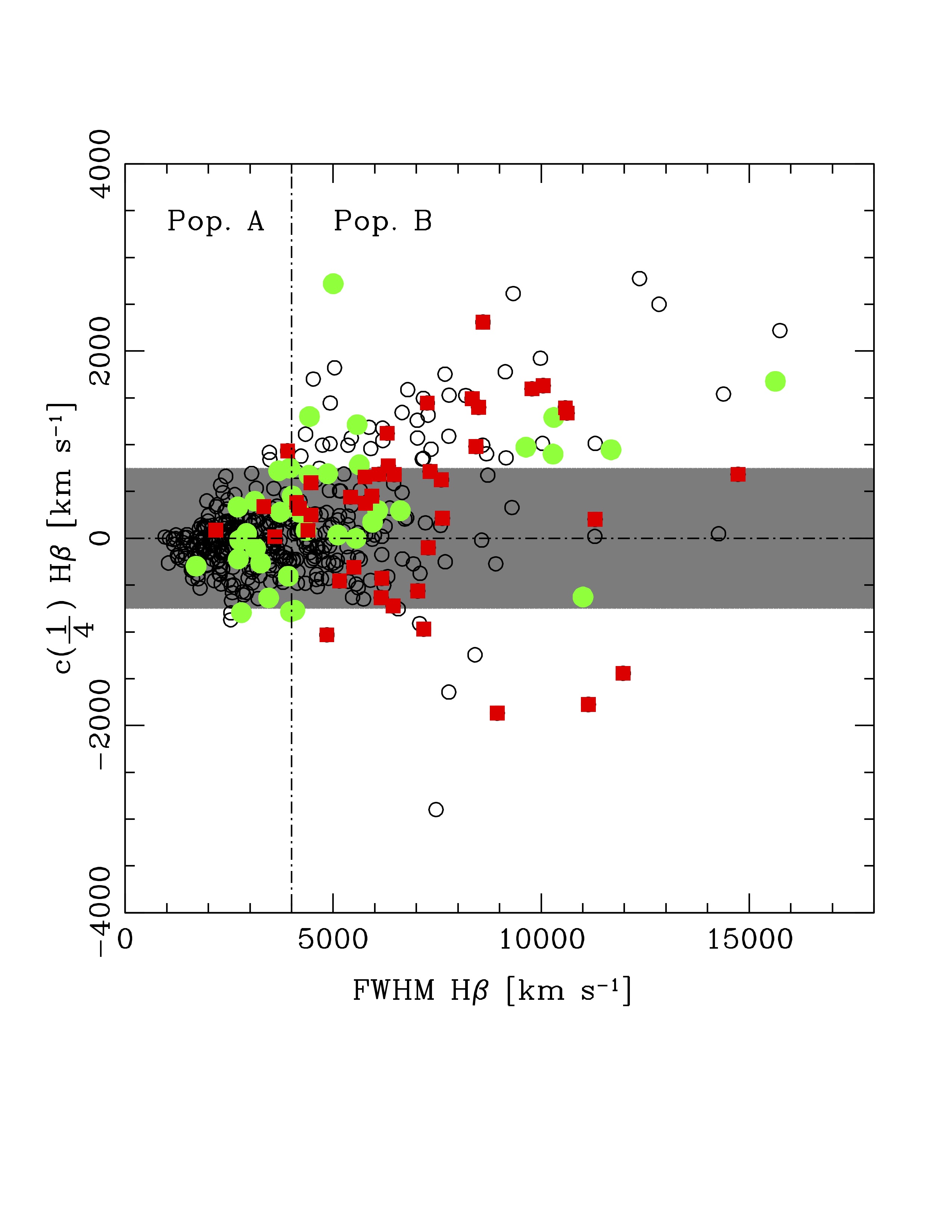}
\vspace{-2cm}
\includegraphics[scale=0.09]{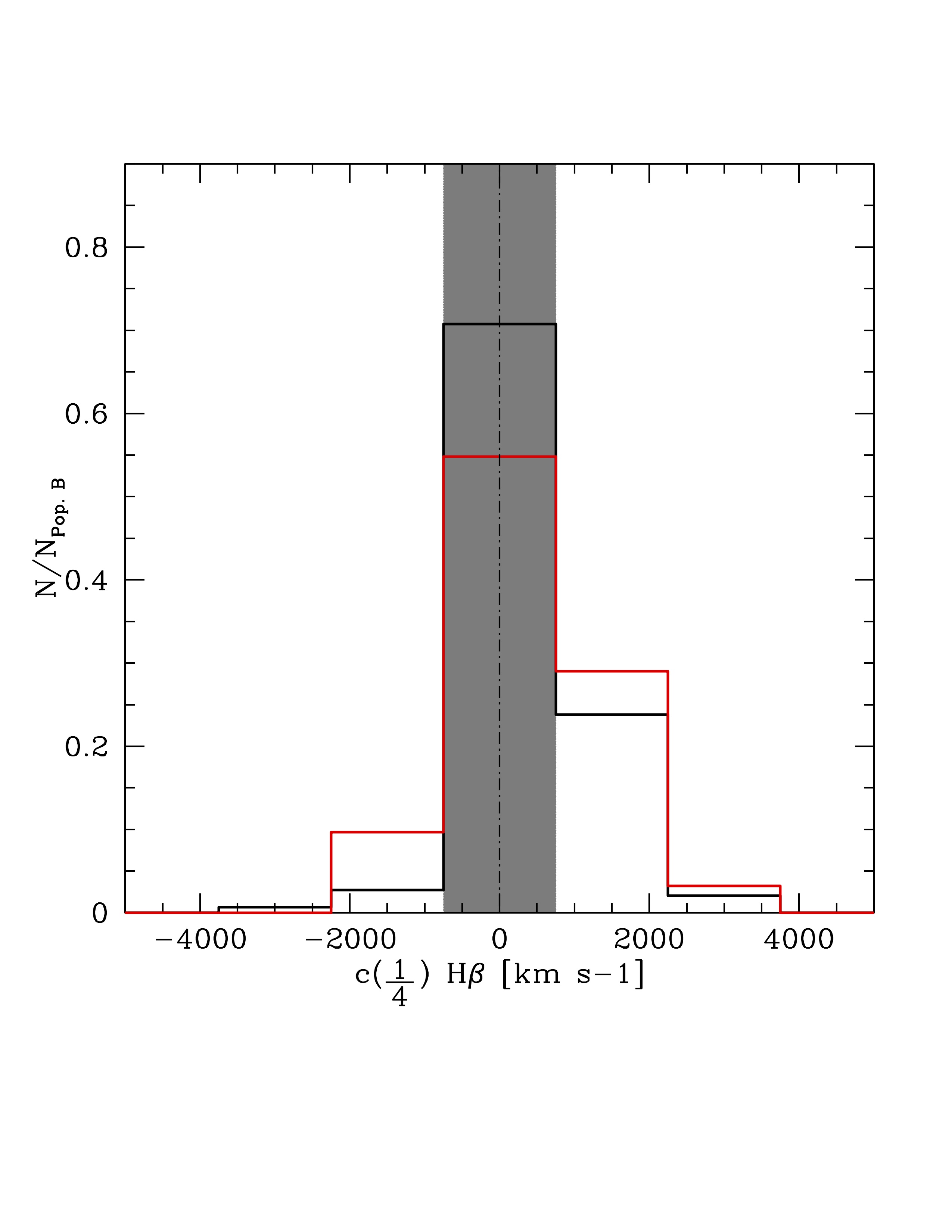}
\caption{Top: FWHM \hb\ vs c(1/4)   parameter  in \kms \ for RQ       (open circles),   core-dominated (green squares) and FRII (red circles) in the \citet{zamfiretal08} sample.  The shaded band width is set  by the uncertainty at a $\pm2 \sigma$\ confidence level in individual AI measurements, as in the previous figure.  Bottom: Histogram of the c(1/4) distribution for Pop. B sources, with bin width  corresponding to the $2 \sigma$\ uncertainty associated with c(1/4), for RQ and RL (red). The restriction to Pop. B is motivated by the paucity of RL sources in Pop. A.   }
\label{fig:par}        
\end{figure}

\begin{figure}[htp!]
\includegraphics[scale=0.09]{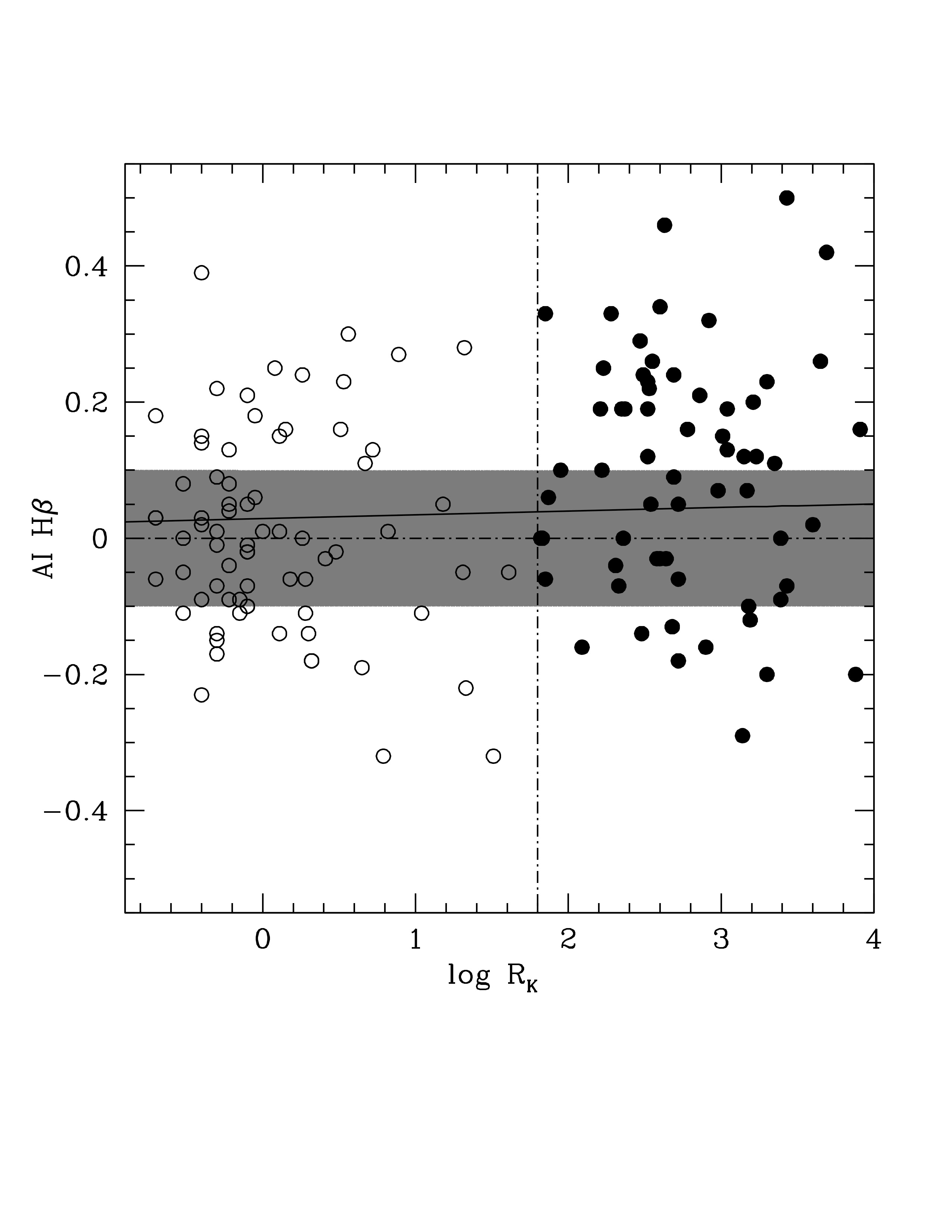}
\vspace{-2cm}
\includegraphics[scale=0.09]{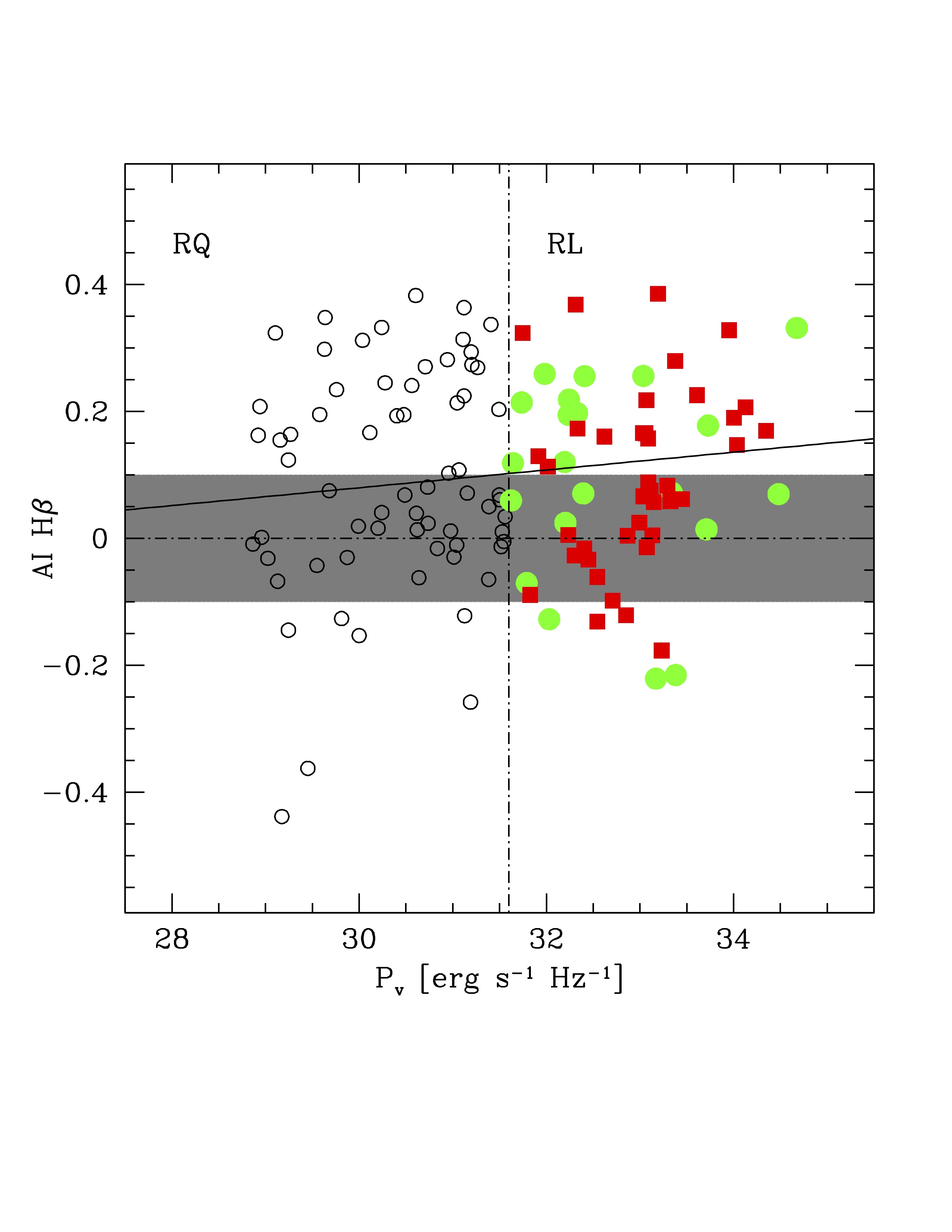}
\caption{Radio-loudness effects  on \hb\ line profiles. Top: \hb\ AI vs. radio-loudness parameter \rk, for radio-detected RQ quasars (open circles) and RL quasars (filled circles) from the sample of \citet{marzianietal03a}. The vertical line shows the nominal boundary for RL objects.   An unweighted lsq fit yields a trend with a slope not significantly different from 0 (straight red line). Bottom: AI vs. radio power at 20cm for Pop. B sources in the sample of \citet{zamfiretal08}. Meaning of symbols is the same as in Fig. \ref{fig:par}.  The slope of the straight line is $0.14 \pm 0.10$, again not significantly different from 0. }
\label{fig:rl}        
\end{figure}

As mentioned, RL effects on LILs can be viewed as second order, if they exist at all. In the plane c(1/4) vs FWHM \hb\ the bulk of the radio loud sources occupies the same area of RQ Pop. B, with some CDs trespassing to the Pop. A side, most likely because of an orientation effect, as they are viewed close to the radio axis (Fig. \ref{fig:par}).   The distribution  of redshifts at 1/4 peak intensity shows no difference for Pop. B RQ and RL, if measurements uncertainties are taken into account (lower panel of Fig. \ref{fig:par}).  There is no correlation with radio-loudness parameters. The panels of Fig. \ref{fig:rl} shows that the \hb\ AI does not depend neither on  \rk\ nor on specific radio power. 

The behaviour of LILs is definitely different from the one of HILs, where the suppression of \civ\ blueshifts occurs mainly in RL sources, making the redward asymmetry associated with the VBC more prominent. In RQ sources, the redward asymmetry is masked by some 
blue shifted emission. The VBLR is believed by some to be a source of strong \civ\ emission \citep{marzianietal10}. The difference between the \hb\ and \civ\ profiles can be explained, at least in part, by adding blue shifted \civ\ emission in RQ sources.

\section{The physical and dynamical  interpretation of line shifts}
\label{physics}
\subsection{Atomic physics}
 
Line broadening and shifts can be produced because of particular physical conditions, for example, for a medium at very high pressure such as in white dwarf atmospheres \citep[][and references 
therein]{halenkaetal15} by quantum mechanical effects \citep{wiesekelleher71}. Such shifts are however rather modest even at the highest density. Generally speaking natural and collisional broadening produce Lorentzian profiles \citep[e.g.,][]{peach81}. We find many Balmer line profiles in Pop. A sources that are Lorentzian-like (to be more precise, they are Voigt due to the convolution with the (Gaussian) instrumental profile). The width is however too large to be produced by collisional effects; Lorentzian profiles are possible also for processes associated with gas motion and geometry.

 In principle, thermal broadening  could  yield a Gaussian line with FWHM$= \sqrt{8 k T_{e} \ln 2 / m_{e}c^{2}} \lambda_{0}$, where $k$ is the Boltzmann constant $T_\mathrm{e}$\ the electron temperature, $m_\mathrm{e}$\ the electron mass, and $\lambda_{0}$\ the line rest frame wavelength. For a typical FWHM \hb\ = 4000 \kms, the electron temperature   needs to be $\sim 10^{9}$ K, too high.

Optical depth effects may also produce  shifts and absorptions dips in the emission profiles. Narrow dips and not widely-spaced double peaks are sometimes observed in \hb\ profiles (typical cases: IC 4329A, 
Akn 120, and OX 169 \citealt{marzianietal92,korista92,halperneracleous00}),  but optical depth effects  
are  usually not considered as a possible cause of the absorption, even if the Balmer line optical 
depth is $\tau \sim 10^{2} \gg 1$. Self-absorption requires that the emitting gas cloud is seen through 
the line of sight \citep{smith80}, which may occur occasionally but not very frequently (a case known 
for many years involves the \ha\ profile of NGC 4151, \citealt{andersonkraft69}). Accordingly, only 
a few cases of \hb\  BAL profiles are known \citep[e.g.,][and references therein]{aokietal06,jietal12}. 
The rarity of these sources is most likely a consequence of the physical conditions needed to produce significant absorption: the maintenance of a large population of HI atoms with electrons excited to level $n = 2$ via \lya\ resonance scattering, so that the $n = 2$\ level can become a pseudo-ground level and \hb\ itself ``behave'' like a resonant line.  

\subsection{Scattering processes}

The BLR is believed to be optically thin for electron scattering \citep{davidsonnetzer79}. However it is still possible that broad line photons are scattered by hot electron surrounding or embedding the line emitting gas \citep{shieldsmckee81}, as in the case of an hot inter cloud medium or of an ionised 
atmosphere above a dense accretion disk. Support for the latter hypothesis has been found by spectropolarimetric analysis of the broad \ha\ profile: in about ten objects, the PA of the polarisation vector rotates across the profile, reaching opposite PA in the red and blue wings \citep{smithetal05}, 
as expected if light is scattered from a rotating disk. Since the emitting region radius is now 
believed to be much smaller than in the past (the inner radius might be less than 100 \rg), it 
is possible that the innermost broad line emission, already associated with gas in rapid motion, is 
further broadened by a screen of hot electrons \citep{laor06}. It is intriguing that \citet{smithetal05} found lower polarisation in the line core. This is consistent with a two component model (i.e., BC + VBC) where the core is emitted further out than the line wings, possibly outside of the screen of scattering material. 

Other scattering processes might be operating as well. Rayleigh scattering could be relevant if there is neutral material 
of high column density, as might be present in an accretion disk or a molecular torus. It could be more important than electron scattering in the UV \citep{gaskellgoosmann13}. Rayleigh scattering could contribute the broadening of the extended line wings 
in \lya\ if \lya\ is emitted above the disk and scattered towards the observer by the dense accretion disk \citep{lee05}. 

Fast (relativistic) electrons moving in a dense med\-ium could be give rise to Cerenkov line-like radiation 
\citep{youetal86,chengetal90}. Densities above \nh$\gtrsim$ $10^{13}$ \cm3 are required for a significant effect. 
This may be the case for an accretion disk as well as for dense gas clumps emitting the LILs in extreme Population A 
sources \citep{negreteetal12}. Shifts on the order of $\sim 1000 $\kms should appear as a net redshift in the Balmer 
and Lyman lines \citep{liuetal14}. 

\subsection{Gravitational redshift}

An early analysis of gravitational redshift in the context of broad line emission was given by \citet{netzer77}. A detailed discussion of gravitational redshift effects on the Balmer line profiles is provided by Bon et al. (submitted for this volume). 
Here we just recall that gravitation can induce a shift \ 
$
\Delta z \approx     \frac{G M_\mathrm{BH}}{R  c^2} =  \frac{1}{2}
\frac{R_{\mathrm{S}}}{R}
$
where  $M_\mathrm{BH}$\ is the black hole mass, and $R_\mathrm{S}$\ the Schwarzschild radius  ($2GM_\mathrm{BH}/c^{2}$). 
This relation is valid with an approximation of 4\%\ down to a 20 Schwarzschild radii. The  total shift of 
line profiles would include the effect of transverse Doppler redshift, so that the expected shift  
is $\Delta z \approx      \frac{3}{4} \frac{R_{\mathrm{S}}}{R}$. Two considerations are in order:
\begin{itemize}
\item the most massive BHs may be associated with the strongest red-ward asymmetries in the \hb\  line \citep{marzianietal09};
\item ray tracing in Kerr metrics produce red wings that reproduce fairly well the observations of observed 
profiles in several high-luminosity sources.
\end{itemize}
Is this enough to justify the claim that grav $z$\ produces the redward asymmetry in Pop. B sources? The answer is "No". The 
problem is that the models assume disk emission with steep emissivity laws that give a significant weight to the 
inner disk regions. Since the radii needed to explain the centroid shifts at 1/4 or the VBC displacements are 
$\sim$ 100 $R_{S}$, there is not likely to be enough cold gas to explain the most luminous sources. A more detailed model 
that includes both kinematics and emitting gas physical conditions (assuming, for instance, that the gas in confined to 
a dense disk) is needed before a conclusion can be reached (Bon et al.).

\subsection{Doppler effect due to an ordered velocity field} 

There are several lines of evidence supporting virial motion as an important source of line broadening:
\begin{itemize}
\item  velocity resolved reverberation mapping studies \citep{koratkargaskell91,grieretal13}  exclude outflows 
as the broadening source at least in the LILs of Population B sources.
\item an anti-correlation between size (derived from reverberation mapping) and line width found for several lines 
in the same sources \citep{petersonwandel99}: a central massive object gravitationally dominates cloud motions, 
with the lines  emitted by ionic species of higher ionization potential due to gas moving faster, and  emitted 
closer to the central massive object. 
\item a fairly symmetric profile is by far the most frequent; apart from the peculiar cases, shifts are usually less 
than 0.1 of the line width at the same fractional height \citep{sulentic89}. 
\item \hb\ and \mgii\ show a trend in line width consistent with the domination of virial motions. \mgii\ is 
preferentially emitted at slightly larger distance than \hb\ from the continuum source \citep{marzianietal13a};
\item there is a gradient in ionization level (approximated in the decomposition BLR/VBLR) that is consistent 
with a virial velocity field.
\end{itemize}

Extreme population A sources (spectral type A3 and A4) often show a blue-shifted excess indicative of radial motion and 
most likely associated with an outflow. This said, the geometry of the gas is not yet clear. The fundamental issue 
of whether the gas is associated with a disk or with a system of clouds is still not settled. Clouds models are still 
formulated at the time of writing perhaps by authors unaware  of the confinement problems that cloud models 
face \citep{shadmehri15}, of which the most serious remains the expected decay time due to the viscous attrition with 
some putative ``inter cloud medium.''  Disk + wind models provide some consistency with ideas about 
Pop. B sources \citep{flohicetal12}, but they still fail to reproduce a typical Pop. B \hb\ profile. At the same time, 
a Lorentzian profile might be reasonably  associated with an extended disk, but other models can also produce Lorentzian or almost Lorentzian  
shapes, most notably a system of clouds under the combined influence of gravity and radiation pressure 
forces \citep[][]{mathews93,netzermarziani10}.

\subsection{Peculiar profiles: Binary BLR, disks and/or outflows?}

\paragraph{Binary BH}  The search for binary black holes is still in its infancy. Apart from a few very special cases 
such as OJ 287 \citep{sillanpaaetal88}, evidence in favour of sub-parsec binaries that could significantly affect the 
broad line profiles is still sparse, and has been hard to find \citep{bonetal12}. Some of the most classical double 
peaked sources (Arp 102B, \citealt{halpernfilippenko88}) show remarkable stability of the peak positions over decades, 
ruling out a binary system with reasonable masses \citep{gezarietal07}. Similar considerations were made for OX 169  \citep{halperneracleous00}. Sub-pc BBH should be numerous at high $z$ but rarer in the local Universe \citep{volonterietal09}. It 
is still unclear whether we should look for a sizeable minority population or for extremely rare event, as coalescence times 
depend on the evolution of the binary orbit eccentricity, and circularised orbit are ``stiff" since they do not yield to 
orbital losses due to gravitational radiation. Profiles with large shifts  are perhaps more promising \citep{borosonlauer09}, and  extensive searches for eventual monitoring are being carried out \citep[e.g.,][]{juetal13}. 

The accretion disk explanation that fits Arp 102B so well fails for other double peaked sources for which a more suitable 
model may be offered by elliptical disks \citep{eracleousetal95}. Accretion disk emission has been postulated for 
the broad line wings \citep{bonetal09a}. Modelling of a disk may require abandoning the weak field treatment in order 
to model the red asymmetry of Balmer line profiles in Pop.  B (Bon et al. submitted).

Single peaked profiles with large shift can be explained by radially moving gas.  However, inferences on the direction of 
radial motions are affected by isotropy of line emission: if the line is anisotropically emitted as expected for the 
Balmer lines in the physical conditions  of the BLR \citep{ferlandetal92}, then a redshift corresponds to inflow (accretion?), 
and a blueshift to outflow (winds?), although it is still possible that we see emission mainly from the shielded face of 
clouds \citep{ferlandetal09}.

\section{How can LIL shifts and other profile properties help  build a physical scenario}
\label{build}

Line profiles of Balmer lines are orientation dependent. This result emerges on a statistical basis by comparing 
the average line widths of lobe-dominated and core-dominated radio-loud sources \citep{willsbrowne86,mcluredunlop02,sulenticetal03}.  Conservation of angular momentum accounts for the the spherical symmetry break that leads to the formation of a Keplerian (or quasi-Keplerian) accretion disk around the black hole. Not surprisingly, a flattened configuration was already ``in place'' in the 
minds of astronomers from heuristic considerations during the 1970s \citep[e.g.,][]{osterbrock78}. However, the viewing angle 
is a parameter that is generally unknown for RQ sources and can be estimated only for superluminal 
radio loud sources on an individual basis \citep{rokakietal03,sulenticetal03}. Approaches using 
polarimetry \citep{afanasievpopovic15} and gravitational redshift (if found appropriate) can in principle lead to an estimate 
of the viewing angle \citep{gavrilovicetal07,kollatschny03}.  A flattened or a ``bowl''-shaped geometry \citep{kollatschnyzetzl11,goadetal12} is   suggested by the dependence of line width on viewing angle. \citet{goadetal12} suggested that line emission 
may occur in the outer region of the accretion disk, where it should merge with the surrounding molecular torus, and were 
able to obtain Lorentzian profiles for the integrated emission from that region. This model is intriguing and may 
well be applicable to extreme Pop. A sources. 

Heuristic considerations and black hole and Eddington ratio estimates suggest that the main driver of the quasar main sequence is Eddington ratio \citep[e.g.,][]{marzianietal01,marzianietal03b,kuraszkiewiczetal09}.  Population A and B are distinct in terms of Eddington ratio: the boundary value (for BH mass of 10$^8$ \msol) is estimated to be \lledd $\approx 0.2 \pm 0.1$ 
\citep{marzianietal03b}. It is interesting to note that this is the limit at which the transition from a geometrically thin to a geometrically thick disk is expected \cite[][and references therein]{abramowiczetal88,franketal02}. So, any model of the 
BLR should be based on this constraint, with models valid for Pop. A including a geometrically thick disk, its anisotropy and 
shadowing effects. A first attempt to include a geometrically thick disk in a BLR model has been carried out by  \citet{wangetal14a}, but the predictions until now contradict the observations.

\section{Key Open issues and conclusions}
\label{open}
 
It is perhaps disappointing  that after 40 years of effort the structure and dynamics of the regions emitting the 
LILs are poorly understood. However, on the bright side,  some key points have been clarified: 

\begin{itemize}
\item we now have a contextualization which allows us to relate, and discriminate between  source classes;
\item a trend in Balmer line shapes is systematic along the 4DE1 sequence;
\item Lorentzian (Pop. A) or double Gaussian (Pop. B) are by far the most frequent \hb\ and \mgii\ profiles;
\item only a minority of sources show very broad profiles that are immediately consistent with accretion disk emission;
\item the wide range of HIL FWHM suggests that black hole mass estimates using this measure as a virial estimator 
must be treated with caution. No model can explain this range invoking only line of sight orientation of a flattened disk.
The rules for mass estimation are likely different for Pop. A and B sources. 
\end{itemize}

The profile differences may be associated with an accretion state transition. 
 
Degeneracy of the regular profile with respect to competing models is still a major problem. Knowing however that \lledd\ is a major driver in HIL profile diversity along the 4DE1 sequence will help building more focused models. 
 
\begin{acknowledgements}
AdO and JS acknowledge the support by the Junta de Andaluc\' \i a through project TIC114,and the Spanish Ministry of Economy and Competitiveness (MINECO) through project AYA2013-42227-P.  
\end{acknowledgements} 
\newpage
\newpage
\pagebreak
%

%


\vfill\eject

\newpage\pagebreak
\bibliographystyle{spr-mp-nameyear-cnd}

\begin{thebibliography}{106}
\ifx \bisbn   \undefined \def \bisbn  #1{ISBN #1}\fi
\ifx \binits  \undefined \def \binits#1{#1} \fi
\ifx \bauthor  \undefined \def \bauthor#1{#1} \fi
\ifx \batitle  \undefined \def \batitle#1{#1} \fi
\ifx \bjtitle  \undefined \def \bjtitle#1{#1}\fi
\ifx \bvolume  \undefined \def \bvolume#1{\textbf{#1}}\fi
\ifx \byear  \undefined \def \byear#1{#1} \fi
\ifx \bissue  \undefined \def \bissue#1{#1} \fi
\ifx \bfpage  \undefined \def \bfpage#1{#1} \fi
\ifx \blpage  \undefined \def \blpage #1{#1} \fi
\ifx \burl  \undefined \def \burl#1{\textsf{#1}} \fi
\ifx \doiurl  \undefined \def \doiurl#1{\textsf{#1}} \fi
\ifx \betal  \undefined \def \betal{\textit{et al.}} \fi
\ifx \binstitute  \undefined \def \binstitute#1{#1} \fi
\ifx \binstitutionaled  \undefined \def \binstitutionaled#1{#1} \fi
\ifx \bctitle  \undefined \def \bctitle#1{#1} \fi
\ifx \beditor  \undefined \def \beditor#1{#1} \fi
\ifx \bpublisher  \undefined \def \bpublisher#1{#1} \fi
\ifx \bbtitle  \undefined \def \bbtitle#1{#1} \fi
\ifx \bedition  \undefined \def \bedition#1{#1} \fi
\ifx \bseriesno  \undefined \def \bseriesno#1{#1} \fi
\ifx \blocation  \undefined \def \blocation#1{#1} \fi
\ifx \bsertitle  \undefined \def \bsertitle#1{#1} \fi
\ifx \bsnm \undefined \def \bsnm#1{#1} \fi
\ifx \bsuffix \undefined \def \bsuffix#1{#1} \fi
\ifx \bparticle \undefined \def \bparticle#1{#1} \fi
\ifx \barticle \undefined \def \barticle#1{#1} \fi
\ifx \bconfdate \undefined \def \bconfdate #1{#1} \fi
\ifx \botherref \undefined \def \botherref #1{#1} \fi
\ifx \url \undefined \def \url#1{\textsf{#1}} \fi
\ifx \bchapter \undefined \def \bchapter#1{#1} \fi
\ifx \bbook \undefined \def \bbook#1{#1} \fi
\ifx \bcomment \undefined \def \bcomment#1{#1} \fi
\ifx \oauthor \undefined \def \oauthor#1{#1} \fi
\ifx \citeauthoryear \undefined \def \citeauthoryear#1{#1} \fi
\ifx \endbibitem  \undefined \def \endbibitem {}\fi
\ifx \bconflocation  \undefined \def \bconflocation#1{#1} \fi
\ifx \arxivurl  \undefined \def \arxivurl#1{\textsf{#1}} \fi

\bibitem[\protect\citeauthoryear{{Abramowicz} et~al.}{1988}]{abramowiczetal88}
\begin{barticle}
\bauthor{\bsnm{{Abramowicz}}, \binits{M.A.}},
\bauthor{\bsnm{{Czerny}}, \binits{B.}},
\bauthor{\bsnm{{Lasota}}, \binits{J.P.}},
\bauthor{\bsnm{{Szuszkiewicz}}, \binits{E.}}:
\bjtitle{\apj}
\bvolume{332},
\bfpage{646}
(\byear{1988}).
doi:\doiurl{10.1086/166683}
\end{barticle}
\endbibitem

\bibitem[\protect\citeauthoryear{{Afanasiev} and
  {Popovi{\'c}}}{2015}]{afanasievpopovic15}
\begin{barticle}
\bauthor{\bsnm{{Afanasiev}}, \binits{V.L.}},
\bauthor{\bsnm{{Popovi{\'c}}}, \binits{L.{\v C}.}}:
\bjtitle{\apjl}
\bvolume{800},
\bfpage{35}
(\byear{2015}).
\arxivurl{1501.07730}.
doi:\doiurl{10.1088/2041-8205/800/2/L35}
\end{barticle}
\endbibitem

\bibitem[\protect\citeauthoryear{{Anderson} and
  {Kraft}}{1969}]{andersonkraft69}
\begin{barticle}
\bauthor{\bsnm{{Anderson}}, \binits{K.S.}},
\bauthor{\bsnm{{Kraft}}, \binits{R.P.}}:
\bjtitle{\apj}
\bvolume{158},
\bfpage{859}
(\byear{1969}).
doi:\doiurl{10.1086/150246}
\end{barticle}
\endbibitem

\bibitem[\protect\citeauthoryear{{Antonucci}}{1993}]{antonucci93}
\begin{barticle}
\bauthor{\bsnm{{Antonucci}}, \binits{R.}}:
\bjtitle{\araa}
\bvolume{31},
\bfpage{473}
(\byear{1993}).
doi:\doiurl{10.1146/annurev.aa.31.090193.002353}
\end{barticle}
\endbibitem

\bibitem[\protect\citeauthoryear{{Antonucci} and
  {Miller}}{1985}]{antonuccimiller85}
\begin{barticle}
\bauthor{\bsnm{{Antonucci}}, \binits{R.R.J.}},
\bauthor{\bsnm{{Miller}}, \binits{J.S.}}:
\bjtitle{\apj}
\bvolume{297},
\bfpage{621}
(\byear{1985}).
doi:\doiurl{10.1086/163559}
\end{barticle}
\endbibitem

\bibitem[\protect\citeauthoryear{{Aoki} et~al.}{2006}]{aokietal06}
\begin{barticle}
\bauthor{\bsnm{{Aoki}}, \binits{K.}},
\bauthor{\bsnm{{Iwata}}, \binits{I.}},
\bauthor{\bsnm{{Ohta}}, \binits{K.}},
\bauthor{\bsnm{{Ando}}, \binits{M.}},
\bauthor{\bsnm{{Akiyama}}, \binits{M.}},
\bauthor{\bsnm{{Tamura}}, \binits{N.}}:
\bjtitle{\apj}
\bvolume{651},
\bfpage{84}
(\byear{2006}).
\arxivurl{arXiv:astro-ph/0607036}.
doi:\doiurl{10.1086/507438}
\end{barticle}
\endbibitem

\bibitem[\protect\citeauthoryear{{Bachev} et~al.}{2004}]{bachevetal04}
\begin{barticle}
\bauthor{\bsnm{{Bachev}}, \binits{R.}},
\bauthor{\bsnm{{Marziani}}, \binits{P.}},
\bauthor{\bsnm{{Sulentic}}, \binits{J.W.}},
\bauthor{\bsnm{{Zamanov}}, \binits{R.}},
\bauthor{\bsnm{{Calvani}}, \binits{M.}},
\bauthor{\bsnm{{Dultzin-Hacyan}}, \binits{D.}}:
\bjtitle{ApJ}
\bvolume{617},
\bfpage{171}
(\byear{2004}).
\arxivurl{arXiv:astro-ph/0408334}.
doi:\doiurl{10.1086/425210}
\end{barticle}
\endbibitem

\bibitem[\protect\citeauthoryear{{Bianchi} et~al.}{2008}]{bianchietal08}
\begin{barticle}
\bauthor{\bsnm{{Bianchi}}, \binits{S.}},
\bauthor{\bsnm{{Corral}}, \binits{A.}},
\bauthor{\bsnm{{Panessa}}, \binits{F.}},
\bauthor{\bsnm{{Barcons}}, \binits{X.}},
\bauthor{\bsnm{{Matt}}, \binits{G.}},
\bauthor{\bsnm{{Bassani}}, \binits{L.}},
\bauthor{\bsnm{{Carrera}}, \binits{F.J.}},
\bauthor{\bsnm{{Jim{\'e}nez-Bail{\'o}n}}, \binits{E.}}:
\bjtitle{\mnras}
\bvolume{385},
\bfpage{195}
(\byear{2008}).
\arxivurl{0710.4226}.
doi:\doiurl{10.1111/j.1365-2966.2007.12625.x}
\end{barticle}
\endbibitem

\bibitem[\protect\citeauthoryear{{Bon} et~al.}{2009}]{bonetal09a}
\begin{barticle}
\bauthor{\bsnm{{Bon}}, \binits{E.}},
\bauthor{\bsnm{{Popovi{\'c}}}, \binits{L.{\v C}.}},
\bauthor{\bsnm{{Gavrilovi{\'c}}}, \binits{N.}},
\bauthor{\bsnm{{Mura}}, \binits{G.L.}},
\bauthor{\bsnm{{Mediavilla}}, \binits{E.}}:
\bjtitle{\mnras}
\bvolume{400},
\bfpage{924}
(\byear{2009}).
\arxivurl{0908.2939}.
doi:\doiurl{10.1111/j.1365-2966.2009.15511.x}
\end{barticle}
\endbibitem

\bibitem[\protect\citeauthoryear{{Bon} et~al.}{2012}]{bonetal12}
\begin{barticle}
\bauthor{\bsnm{{Bon}}, \binits{E.}},
\bauthor{\bsnm{{Jovanovi{\'c}}}, \binits{P.}},
\bauthor{\bsnm{{Marziani}}, \binits{P.}},
\bauthor{\bsnm{{Shapovalova}}, \binits{A.I.}},
\bauthor{\bsnm{{Bon}}, \binits{N.}},
\bauthor{\bsnm{{Borka Jovanovi{\'c}}}, \binits{V.}},
\bauthor{\bsnm{{Borka}}, \binits{D.}},
\bauthor{\bsnm{{Sulentic}}, \binits{J.}},
\bauthor{\bsnm{{Popovi{\'c}}}, \binits{L.{\v C}.}}:
\bjtitle{\apj}
\bvolume{759},
\bfpage{118}
(\byear{2012}).
\arxivurl{1209.4524}.
doi:\doiurl{10.1088/0004-637X/759/2/118}
\end{barticle}
\endbibitem

\bibitem[\protect\citeauthoryear{{Boroson} and {Green}}{1992}]{borosongreen92}
\begin{barticle}
\bauthor{\bsnm{{Boroson}}, \binits{T.A.}},
\bauthor{\bsnm{{Green}}, \binits{R.F.}}:
\bjtitle{ApJS}
\bvolume{80},
\bfpage{109}
(\byear{1992}).
doi:\doiurl{10.10\-86\-/\-19\-1661}
\end{barticle}
\endbibitem

\bibitem[\protect\citeauthoryear{{Boroson} and {Lauer}}{2009}]{borosonlauer09}
\begin{barticle}
\bauthor{\bsnm{{Boroson}}, \binits{T.A.}},
\bauthor{\bsnm{{Lauer}}, \binits{T.R.}}:
\bjtitle{Nature}
\bvolume{458},
\bfpage{53}
(\byear{2009}).
\arxivurl{0901.3779}.
doi:\doiurl{10.1038/nature07779}
\end{barticle}
\endbibitem

\bibitem[\protect\citeauthoryear{{Chen} et~al.}{1989}]{chenetal89}
\begin{barticle}
\bauthor{\bsnm{{Chen}}, \binits{K.}},
\bauthor{\bsnm{{Halpern}}, \binits{J.P.}},
\bauthor{\bsnm{{Filippenko}}, \binits{A.V.}}:
\bjtitle{\apj}
\bvolume{339},
\bfpage{742}
(\byear{1989}).
doi:\doiurl{10.1086/167332}
\end{barticle}
\endbibitem

\bibitem[\protect\citeauthoryear{{Cheng} et~al.}{1990}]{chengetal90}
\begin{barticle}
\bauthor{\bsnm{{Cheng}}, \binits{F.H.}},
\bauthor{\bsnm{{You}}, \binits{J.H.}},
\bauthor{\bsnm{{Yan}}, \binits{M.}}:
\bjtitle{\apj}
\bvolume{358},
\bfpage{18}
(\byear{1990}).
doi:\doiurl{10.1086/168958}
\end{barticle}
\endbibitem

\bibitem[\protect\citeauthoryear{{Collin-Souffrin}
  et~al.}{1988}]{collinsouffrinetal88}
\begin{barticle}
\bauthor{\bsnm{{Collin-Souffrin}}, \binits{S.}},
\bauthor{\bsnm{{Dyson}}, \binits{J.E.}},
\bauthor{\bsnm{{McDowell}}, \binits{J.C.}},
\bauthor{\bsnm{{Perry}}, \binits{J.J.}}:
\bjtitle{MNRAS}
\bvolume{232},
\bfpage{539}
(\byear{1988})
\end{barticle}
\endbibitem

\bibitem[\protect\citeauthoryear{{Corbin} and
  {Boroson}}{1996}]{corbinboroson96}
\begin{barticle}
\bauthor{\bsnm{{Corbin}}, \binits{M.R.}},
\bauthor{\bsnm{{Boroson}}, \binits{T.A.}}:
\bjtitle{\apjs}
\bvolume{107},
\bfpage{69}
(\byear{1996}).
doi:\doiurl{10.1086/192355}
\end{barticle}
\endbibitem

\bibitem[\protect\citeauthoryear{{Davidson} and
  {Netzer}}{1979}]{davidsonnetzer79}
\begin{barticle}
\bauthor{\bsnm{{Davidson}}, \binits{K.}},
\bauthor{\bsnm{{Netzer}}, \binits{H.}}:
\bjtitle{Reviews of Modern Physics}
\bvolume{51},
\bfpage{715}
(\byear{1979}).
doi:\doiurl{10.1103/RevModPhys.51.715}
\end{barticle}
\endbibitem

\bibitem[\protect\citeauthoryear{{Done} et~al.}{2012}]{doneetal12}
\begin{barticle}
\bauthor{\bsnm{{Done}}, \binits{C.}},
\bauthor{\bsnm{{Davis}}, \binits{S.W.}},
\bauthor{\bsnm{{Jin}}, \binits{C.}},
\bauthor{\bsnm{{Blaes}}, \binits{O.}},
\bauthor{\bsnm{{Ward}}, \binits{M.}}:
\bjtitle{\mnras}
\bvolume{420},
\bfpage{1848}
(\byear{2012}).
\arxivurl{1107.5429}.
doi:\doiurl{10.1111/j.1365-2966.2011.19779.x}
\end{barticle}
\endbibitem

\bibitem[\protect\citeauthoryear{{D'Onofrio} et~al.}{2012}]{donofrioetal12}
\begin{bbook}
\beditor{\bsnm{{D'Onofrio}}, \binits{M.}},
\beditor{\bsnm{{Marziani}}, \binits{P.}},
\beditor{\bsnm{{ Sulentic}}, \binits{J.W.}} (eds.):
\bbtitle{Fifty Years of Quasars From Early Observations and Ideas to Future
  Research}.
\bsertitle{Astrophysics and Space Science Library},
vol. \bseriesno{386}.
\bpublisher{Springer Verlag, Berlin-Heidelberg}, \blocation{???}
(\byear{2012})
\end{bbook}
\endbibitem

\bibitem[\protect\citeauthoryear{{Elvis}}{2000}]{elvis00}
\begin{barticle}
\bauthor{\bsnm{{Elvis}}, \binits{M.}}:
\bjtitle{\apj}
\bvolume{545},
\bfpage{63}
(\byear{2000}).
\arxivurl{arXiv:astro-ph/0008064}.
doi:\doiurl{10.1086/317778}
\end{barticle}
\endbibitem

\bibitem[\protect\citeauthoryear{{Eracleous} et~al.}{1995}]{eracleousetal95}
\begin{barticle}
\bauthor{\bsnm{{Eracleous}}, \binits{M.}},
\bauthor{\bsnm{{Livio}}, \binits{M.}},
\bauthor{\bsnm{{Halpern}}, \binits{J.P.}},
\bauthor{\bsnm{{Storchi-Bergmann}}, \binits{T.}}:
\bjtitle{\apj}
\bvolume{438},
\bfpage{610}
(\byear{1995}).
doi:\doiurl{10.1086/175104}
\end{barticle}
\endbibitem

\bibitem[\protect\citeauthoryear{{Ferland} et~al.}{1992}]{ferlandetal92}
\begin{barticle}
\bauthor{\bsnm{{Ferland}}, \binits{G.J.}},
\bauthor{\bsnm{{Peterson}}, \binits{B.M.}},
\bauthor{\bsnm{{Horne}}, \binits{K.}},
\bauthor{\bsnm{{Welsh}}, \binits{W.F.}},
\bauthor{\bsnm{{Nahar}}, \binits{S.N.}}:
\bjtitle{\apj}
\bvolume{387},
\bfpage{95}
(\byear{1992}).
doi:\doiurl{10.1086/\-171063}
\end{barticle}
\endbibitem

\bibitem[\protect\citeauthoryear{{Ferland} et~al.}{2009}]{ferlandetal09}
\begin{barticle}
\bauthor{\bsnm{{Ferland}}, \binits{G.J.}},
\bauthor{\bsnm{{Hu}}, \binits{C.}},
\bauthor{\bsnm{{Wang}}, \binits{J.}},
\bauthor{\bsnm{{Baldwin}}, \binits{J.A.}},
\bauthor{\bsnm{{Porter}}, \binits{R.L.}},
\bauthor{\bsnm{{van Hoof}}, \binits{P.A.M.}},
\bauthor{\bsnm{{Williams}}, \binits{R.J.R.}}:
\bjtitle{\apjl}
\bvolume{707},
\bfpage{82}
(\byear{2009}).
\arxivurl{0911.1173}.
doi:\doiurl{10.1088/0004-637X/707/1/L82}
\end{barticle}
\endbibitem

\bibitem[\protect\citeauthoryear{{Flohic} et~al.}{2012}]{flohicetal12}
\begin{barticle}
\bauthor{\bsnm{{Flohic}}, \binits{H.M.L.G.}},
\bauthor{\bsnm{{Eracleous}}, \binits{M.}},
\bauthor{\bsnm{{Bogdanovi{\'c}}}, \binits{T.}}:
\bjtitle{ApJ}
\bvolume{753},
\bfpage{133}
(\byear{2012}).
doi:\doiurl{10.1088/0004-637X/753/2/133}
\end{barticle}
\endbibitem

\bibitem[\protect\citeauthoryear{{Frank} et~al.}{2002}]{franketal02}
\begin{bbook}
\bauthor{\bsnm{{Frank}}, \binits{J.}},
\bauthor{\bsnm{{King}}, \binits{A.}},
\bauthor{\bsnm{{Raine}}, \binits{D.J.}}:
\bbtitle{Accretion Power in Astrophysics: Third Edition},
\bedition{Iii edition} edn.
\bpublisher{Cambridge University Press}, \blocation{???}
(\byear{2002})
\end{bbook}
\endbibitem

\bibitem[\protect\citeauthoryear{{Gaskell}}{1982}]{gaskell82}
\begin{barticle}
\bauthor{\bsnm{{Gaskell}}, \binits{C.M.}}:
\bjtitle{ApJ}
\bvolume{263},
\bfpage{79}
(\byear{1982}).
doi:\doiurl{10.1086/160481}
\end{barticle}
\endbibitem

\bibitem[\protect\citeauthoryear{{Gaskell} and
  {Goosmann}}{2013}]{gaskellgoosmann13}
\begin{barticle}
\bauthor{\bsnm{{Gaskell}}, \binits{C.M.}},
\bauthor{\bsnm{{Goosmann}}, \binits{R.W.}}:
\bjtitle{\apj}
\bvolume{769},
\bfpage{30}
(\byear{2013}).
\arxivurl{0805.4258}.
doi:\doiurl{10.1088/0004-637X/769/1/30}
\end{barticle}
\endbibitem

\bibitem[\protect\citeauthoryear{{Gavrilovi{\'c}}
  et~al.}{2007}]{gavrilovicetal07}
\begin{bchapter}
\bauthor{\bsnm{{Gavrilovi{\'c}}}, \binits{N.}},
\bauthor{\bsnm{{Popovi{\'c}}}, \binits{L.{\v C}.}},
\bauthor{\bsnm{{Kollatschny}}, \binits{W.}}:
In: \beditor{\bsnm{{Karas}}, \binits{V.}},
\beditor{\bsnm{{Matt}}, \binits{G.}} (eds.)
\bbtitle{IAU Symposium}.
\bsertitle{IAU Symposium},
vol. \bseriesno{238},
p. \bfpage{369}
(\byear{2007}).
doi:\doiurl{10.1017/S1743921307005492}
\end{bchapter}
\endbibitem

\bibitem[\protect\citeauthoryear{{Gezari} et~al.}{2007}]{gezarietal07}
\begin{barticle}
\bauthor{\bsnm{{Gezari}}, \binits{S.}},
\bauthor{\bsnm{{Halpern}}, \binits{J.P.}},
\bauthor{\bsnm{{Eracleous}}, \binits{M.}}:
\bjtitle{\apjs}
\bvolume{169},
\bfpage{167}
(\byear{2007}).
\arxivurl{arXiv:astro-ph/0702594}.
doi:\doiurl{10.1086/511032}
\end{barticle}
\endbibitem

\bibitem[\protect\citeauthoryear{{Goad} and {Korista}}{2014}]{goadkorista14}
\begin{barticle}
\bauthor{\bsnm{{Goad}}, \binits{M.R.}},
\bauthor{\bsnm{{Korista}}, \binits{K.T.}}:
\bjtitle{\mnras}
\bvolume{444},
\bfpage{43}
(\byear{2014}).
\arxivurl{1407.5004}.
doi:\doiurl{10.1093/mnras/stu1456}
\end{barticle}
\endbibitem

\bibitem[\protect\citeauthoryear{{Goad} et~al.}{2012}]{goadetal12}
\begin{barticle}
\bauthor{\bsnm{{Goad}}, \binits{M.R.}},
\bauthor{\bsnm{{Korista}}, \binits{K.T.}},
\bauthor{\bsnm{{Ruff}}, \binits{A.J.}}:
\bjtitle{\mnras}
\bvolume{426},
\bfpage{3086}
(\byear{2012}).
\arxivurl{1207.6339}.
doi:\doiurl{10.1111/\-j.1365\--2966.2012.21808.x}
\end{barticle}
\endbibitem

\bibitem[\protect\citeauthoryear{{Greenstein} and
  {Schmidt}}{1964}]{greensteinschmidt64}
\begin{barticle}
\bauthor{\bsnm{{Greenstein}}, \binits{J.L.}},
\bauthor{\bsnm{{Schmidt}}, \binits{M.}}:
\bjtitle{\apj}
\bvolume{140},
\bfpage{1}
(\byear{1964}).
doi:\doiurl{10.1086/147889}
\end{barticle}
\endbibitem

\bibitem[\protect\citeauthoryear{{Grier} et~al.}{2013}]{grieretal13}
\begin{barticle}
\bauthor{\bsnm{{Grier}}, \binits{C.J.}},
\bauthor{\bsnm{{Peterson}}, \binits{B.M.}},
\bauthor{\bsnm{{Horne}}, \binits{K.}},
\bauthor{\bsnm{{Bentz}}, \binits{M.C.}},
\bauthor{\bsnm{{Pogge}}, \binits{R.W.}},
\bauthor{\bsnm{{Denney}}, \binits{K.D.}},
\bauthor{\bsnm{{De Rosa}}, \binits{G.}},
\bauthor{\bsnm{{Martini}}, \binits{P.}},
\bauthor{\bsnm{{Kochanek}}, \binits{C.S.}},
\bauthor{\bsnm{{Zu}}, \binits{Y.}},
\bauthor{\bsnm{{Shappee}}, \binits{B.}},
\bauthor{\bsnm{{Siverd}}, \binits{R.}},
\bauthor{\bsnm{{Beatty}}, \binits{T.G.}},
\bauthor{\bsnm{{Sergeev}}, \binits{S.G.}},
\bauthor{\bsnm{{Kaspi}}, \binits{S.}},
\bauthor{\bsnm{{Araya Salvo}}, \binits{C.}},
\bauthor{\bsnm{{Bird}}, \binits{J.C.}},
\bauthor{\bsnm{{Bord}}, \binits{D.J.}},
\bauthor{\bsnm{{Borman}}, \binits{G.A.}},
\bauthor{\bsnm{{Che}}, \binits{X.}},
\bauthor{\bsnm{{Chen}}, \binits{C.}},
\bauthor{\bsnm{{Cohen}}, \binits{S.A.}},
\bauthor{\bsnm{{Dietrich}}, \binits{M.}},
\bauthor{\bsnm{{Doroshenko}}, \binits{V.T.}},
\bauthor{\bsnm{{Efimov}}, \binits{Y.S.}},
\bauthor{\bsnm{{Free}}, \binits{N.}},
\bauthor{\bsnm{{Ginsburg}}, \binits{I.}},
\bauthor{\bsnm{{Henderson}}, \binits{C.B.}},
\bauthor{\bsnm{{King}}, \binits{A.L.}},
\bauthor{\bsnm{{Mogren}}, \binits{K.}},
\bauthor{\bsnm{{Molina}}, \binits{M.}},
\bauthor{\bsnm{{Mosquera}}, \binits{A.M.}},
\bauthor{\bsnm{{Nazarov}}, \binits{S.V.}},
\bauthor{\bsnm{{Okhmat}}, \binits{D.N.}},
\bauthor{\bsnm{{Pejcha}}, \binits{O.}},
\bauthor{\bsnm{{Rafter}}, \binits{S.}},
\bauthor{\bsnm{{Shields}}, \binits{J.C.}},
\bauthor{\bsnm{{Skowron}}, \binits{J.}},
\bauthor{\bsnm{{Szczygiel}}, \binits{D.M.}},
\bauthor{\bsnm{{Valluri}}, \binits{M.}},
\bauthor{\bsnm{{van Saders}}, \binits{J.L.}}:
\bjtitle{\apj}
\bvolume{764},
\bfpage{47}
(\byear{2013}).
\arxivurl{1210.2397}.
doi:\doiurl{10.1088/0004-637X/764/1/47}
\end{barticle}
\endbibitem

\bibitem[\protect\citeauthoryear{{Haardt} and
  {Maraschi}}{1991}]{haardtmaraschi91}
\begin{barticle}
\bauthor{\bsnm{{Haardt}}, \binits{F.}},
\bauthor{\bsnm{{Maraschi}}, \binits{L.}}:
\bjtitle{\apjl}
\bvolume{380},
\bfpage{51}
(\byear{1991}).
doi:\doiurl{10.1086/186171}
\end{barticle}
\endbibitem

\bibitem[\protect\citeauthoryear{{Halenka} et~al.}{2015}]{halenkaetal15}
\begin{barticle}
\bauthor{\bsnm{{Halenka}}, \binits{J.}},
\bauthor{\bsnm{{Olchawa}}, \binits{W.}},
\bauthor{\bsnm{{Madej}}, \binits{J.}},
\bauthor{\bsnm{{Grabowski}}, \binits{B.}}:
\bjtitle{\apj}
\bvolume{808},
\bfpage{131}
(\byear{2015}).
\arxivurl{1506.04064}.
doi:\doiurl{10.1088/0004-637X/808/2/131}
\end{barticle}
\endbibitem

\bibitem[\protect\citeauthoryear{{Halpern} and
  {Eracleous}}{2000}]{halperneracleous00}
\begin{barticle}
\bauthor{\bsnm{{Halpern}}, \binits{J.P.}},
\bauthor{\bsnm{{Eracleous}}, \binits{M.}}:
\bjtitle{\apj}
\bvolume{531},
\bfpage{647}
(\byear{2000}).
\arxivurl{astro-ph/9910118}.
doi:\doiurl{10.1086/308516}
\end{barticle}
\endbibitem

\bibitem[\protect\citeauthoryear{{Halpern} and
  {Filippenko}}{1988}]{halpernfilippenko88}
\begin{barticle}
\bauthor{\bsnm{{Halpern}}, \binits{J.P.}},
\bauthor{\bsnm{{Filippenko}}, \binits{A.V.}}:
\bjtitle{Nature}
\bvolume{331},
\bfpage{46}
(\byear{1988}).
doi:\doiurl{10.1038/331046a0}
\end{barticle}
\endbibitem

\bibitem[\protect\citeauthoryear{{Hu} et~al.}{2008}]{huetal08}
\begin{barticle}
\bauthor{\bsnm{{Hu}}, \binits{C.}},
\bauthor{\bsnm{{Wang}}, \binits{J.-M.}},
\bauthor{\bsnm{{Ho}}, \binits{L.C.}},
\bauthor{\bsnm{{Chen}}, \binits{Y.-M.}},
\bauthor{\bsnm{{Bian}}, \binits{W.-H.}},
\bauthor{\bsnm{{Xue}}, \binits{S.-J.}}:
\bjtitle{ApJL}
\bvolume{683},
\bfpage{115}
(\byear{2008}).
\arxivurl{0807.2060}.
doi:\doiurl{10.1086/591848}
\end{barticle}
\endbibitem

\bibitem[\protect\citeauthoryear{{Ji} et~al.}{2012}]{jietal12}
\begin{barticle}
\bauthor{\bsnm{{Ji}}, \binits{T.}},
\bauthor{\bsnm{{Wang}}, \binits{T.-G.}},
\bauthor{\bsnm{{Zhou}}, \binits{H.-Y.}},
\bauthor{\bsnm{{Wang}}, \binits{H.-Y.}}:
\bjtitle{Research in Astronomy and Astrophysics}
\bvolume{12},
\bfpage{369}
(\byear{2012}).
\arxivurl{1201.1054}.
doi:\doiurl{10.1088/1674-4527/12/4/002}
\end{barticle}
\endbibitem

\bibitem[\protect\citeauthoryear{{Jin} et~al.}{2009}]{jinetal09}
\begin{barticle}
\bauthor{\bsnm{{Jin}}, \binits{C.}},
\bauthor{\bsnm{{Done}}, \binits{C.}},
\bauthor{\bsnm{{Ward}}, \binits{M.}},
\bauthor{\bsnm{{Gierli{\'n}ski}}, \binits{M.}},
\bauthor{\bsnm{{Mullaney}}, \binits{J.}}:
\bjtitle{\mnras}
\bvolume{398},
\bfpage{16}
(\byear{2009}).
\arxivurl{0903.4698}.
doi:\doiurl{10.1111/j.1745-3933.2009.00697.x}
\end{barticle}
\endbibitem

\bibitem[\protect\citeauthoryear{{Ju} et~al.}{2013}]{juetal13}
\begin{barticle}
\bauthor{\bsnm{{Ju}}, \binits{W.}},
\bauthor{\bsnm{{Greene}}, \binits{J.E.}},
\bauthor{\bsnm{{Rafikov}}, \binits{R.R.}},
\bauthor{\bsnm{{Bickerton}}, \binits{S.J.}},
\bauthor{\bsnm{{Badenes}}, \binits{C.}}:
\bjtitle{\apj}
\bvolume{777},
\bfpage{44}
(\byear{2013}).
\arxivurl{1306.4987}.
doi:\doiurl{10.1088/0004-637X/777/1/44}
\end{barticle}
\endbibitem

\bibitem[\protect\citeauthoryear{{Kollatschny}}{2003}]{kollatschny03}
\begin{barticle}
\bauthor{\bsnm{{Kollatschny}}, \binits{W.}}:
\bjtitle{\aap}
\bvolume{407},
\bfpage{461}
(\byear{2003}).
\arxivurl{astro-ph/0306389}.
doi:\doiurl{10.1051/0004-6361:20030928}
\end{barticle}
\endbibitem

\bibitem[\protect\citeauthoryear{{Kollatschny} and
  {Zetzl}}{2011}]{kollatschnyzetzl11}
\begin{barticle}
\bauthor{\bsnm{{Kollatschny}}, \binits{W.}},
\bauthor{\bsnm{{Zetzl}}, \binits{M.}}:
\bjtitle{\nat}
\bvolume{470},
\bfpage{366}
(\byear{2011}).
doi:\doiurl{10.1038/nature09761}
\end{barticle}
\endbibitem

\bibitem[\protect\citeauthoryear{{Koratkar} and
  {Gaskell}}{1991}]{koratkargaskell91}
\begin{barticle}
\bauthor{\bsnm{{Koratkar}}, \binits{A.P.}},
\bauthor{\bsnm{{Gaskell}}, \binits{C.M.}}:
\bjtitle{\apjs}
\bvolume{75},
\bfpage{719}
(\byear{1991}).
doi:\doiurl{10.1086/191547}
\end{barticle}
\endbibitem

\bibitem[\protect\citeauthoryear{{Korista}}{1992}]{korista92}
\begin{barticle}
\bauthor{\bsnm{{Korista}}, \binits{K.T.}}:
\bjtitle{\apjs}
\bvolume{79},
\bfpage{285}
(\byear{1992}).
doi:\doiurl{10.1086/191654}
\end{barticle}
\endbibitem

\bibitem[\protect\citeauthoryear{{Kuraszkiewicz}
  et~al.}{2009}]{kuraszkiewiczetal09}
\begin{barticle}
\bauthor{\bsnm{{Kuraszkiewicz}}, \binits{J.}},
\bauthor{\bsnm{{Wilkes}}, \binits{B.J.}},
\bauthor{\bsnm{{Schmidt}}, \binits{G.}},
\bauthor{\bsnm{{Smith}}, \binits{P.S.}},
\bauthor{\bsnm{{Cutri}}, \binits{R.}},
\bauthor{\bsnm{{Czerny}}, \binits{B.}}:
\bjtitle{\apj}
\bvolume{692},
\bfpage{1180}
(\byear{2009}).
doi:\doiurl{10.1088/0004-637X/692/2/1180}
\end{barticle}
\endbibitem

\bibitem[\protect\citeauthoryear{{Laor}}{2003}]{laor03}
\begin{barticle}
\bauthor{\bsnm{{Laor}}, \binits{A.}}:
\bjtitle{\apj}
\bvolume{590},
\bfpage{86}
(\byear{2003}).
\arxivurl{astro-ph/0302541}.
doi:\doiurl{10.1086/375008}
\end{barticle}
\endbibitem

\bibitem[\protect\citeauthoryear{{Laor}}{2006}]{laor06}
\begin{barticle}
\bauthor{\bsnm{{Laor}}, \binits{A.}}:
\bjtitle{\apj}
\bvolume{643},
\bfpage{112}
(\byear{2006}).
\arxivurl{astro-ph/0601688}.
doi:\doiurl{10.1086/502798}
\end{barticle}
\endbibitem

\bibitem[\protect\citeauthoryear{{Lee}}{2005}]{lee05}
\begin{bchapter}
\bauthor{\bsnm{{Lee}}, \binits{H.-W.}}:
In: \beditor{\bsnm{{Adamson}}, \binits{A.}},
\beditor{\bsnm{{Aspin}}, \binits{C.}},
\beditor{\bsnm{{Davis}}, \binits{C.}},
\beditor{\bsnm{{Fujiyoshi}}, \binits{T.}} (eds.)
\bbtitle{Astronomical Polarimetry: Current Status and Future Directions}.
\bsertitle{Astronomical Society of the Pacific Conference Series},
vol. \bseriesno{343},
p. \bfpage{441}
(\byear{2005})
\end{bchapter}
\endbibitem

\bibitem[\protect\citeauthoryear{{Leighly} and {Moore}}{2004}]{leighlymoore04}
\begin{barticle}
\bauthor{\bsnm{{Leighly}}, \binits{K.M.}},
\bauthor{\bsnm{{Moore}}, \binits{J.R.}}:
\bjtitle{\apj}
\bvolume{611},
\bfpage{107}
(\byear{2004}).
\arxivurl{arXiv:astro-ph/0402453}.
doi:\doiurl{10.1086/422088}
\end{barticle}
\endbibitem

\bibitem[\protect\citeauthoryear{{Liu} et~al.}{2014}]{liuetal14}
\begin{barticle}
\bauthor{\bsnm{{Liu}}, \binits{D.B.}},
\bauthor{\bsnm{{Chen}}, \binits{W.P.}},
\bauthor{\bsnm{{You}}, \binits{J.H.}},
\bauthor{\bsnm{{Chen}}, \binits{L.}}:
\bjtitle{\apj}
\bvolume{780},
\bfpage{89}
(\byear{2014}).
doi:\doiurl{10.1088/0004-637X/780/1/89}
\end{barticle}
\endbibitem

\bibitem[\protect\citeauthoryear{{ Mart{\'{\i}}nez-Aldama}
  et~al.}{2015}]{martinez-aldamaetal15}
\begin{barticle}
\bauthor{\bsnm{{Loli Mart{\'{\i}}nez-Aldama}}, \binits{M. L.}},
\bauthor{\bsnm{{Dultzin}}, \binits{D.}},
\bauthor{\bsnm{{Marziani}}, \binits{P.}},
\bauthor{\bsnm{{Sulentic}}, \binits{J.W.}},
\bauthor{\bsnm{{Bressan}}, \binits{A.}},
\bauthor{\bsnm{{Chen}}, \binits{Y.}},
\bauthor{\bsnm{{Stirpe}}, \binits{G.M.}}:
\bjtitle{\apjs}
\bvolume{217},
\bfpage{3}
(\byear{2015}).
\arxivurl{1501.04718}.
doi:\doiurl{10.1088/0067-0049/217/1/3}
\end{barticle}
\endbibitem

\bibitem[\protect\citeauthoryear{{Marziani} et~al.}{1992}]{marzianietal92}
\begin{barticle}
\bauthor{\bsnm{{Marziani}}, \binits{P.}},
\bauthor{\bsnm{{Calvani}}, \binits{M.}},
\bauthor{\bsnm{{Sulentic}}, \binits{J.W.}}:
\bjtitle{ApJ}
\bvolume{393},
\bfpage{658}
(\byear{1992}).
doi:\doiurl{10.1086/171533}
\end{barticle}
\endbibitem

\bibitem[\protect\citeauthoryear{{Marziani} et~al.}{2006}]{marzianietal06}
\begin{bbook}
\bauthor{\bsnm{{Marziani}}, \binits{P.}},
\bauthor{\bsnm{{Dultzin-Hacyan}}, \binits{D.}},
\bauthor{\bsnm{{Sulentic}}, \binits{J.W.}}:
In: \beditor{\bsnm{{Kreitler}}, \binits{P.V.}} (ed.)
\bbtitle{{Accretion onto Supermassive Black Holes in Quasars: Learning from
  Optical/UV Observations}}
vol. \bseriesno{New Developments in Black Hole Research},
p. \bfpage{123}.
\bpublisher{Nova Press, New York}, \blocation{???}
(\byear{2006})
\end{bbook}
\endbibitem

\bibitem[\protect\citeauthoryear{{Marziani} et~al.}{1993}]{marzianietal93}
\begin{barticle}
\bauthor{\bsnm{{Marziani}}, \binits{P.}},
\bauthor{\bsnm{{Sulentic}}, \binits{J.W.}},
\bauthor{\bsnm{{Calvani}}, \binits{M.}},
\bauthor{\bsnm{{Perez}}, \binits{E.}},
\bauthor{\bsnm{{Moles}}, \binits{M.}},
\bauthor{\bsnm{{Penston}}, \binits{M.V.}}:
\bjtitle{ApJ}
\bvolume{410},
\bfpage{56}
(\byear{1993}).
\arxivurl{arXiv:astro-ph/9301001}.
doi:\doiurl{10.1086/172724}
\end{barticle}
\endbibitem

\bibitem[\protect\citeauthoryear{{Marziani} et~al.}{1996}]{marzianietal96}
\begin{barticle}
\bauthor{\bsnm{{Marziani}}, \binits{P.}},
\bauthor{\bsnm{{Sulentic}}, \binits{J.W.}},
\bauthor{\bsnm{{Dultzin-Hacyan}}, \binits{D.}},
\bauthor{\bsnm{{Calvani}}, \binits{M.}},
\bauthor{\bsnm{{Moles}}, \binits{M.}}:
\bjtitle{ApJS}
\bvolume{104},
\bfpage{37}
(\byear{1996}).
doi:\doiurl{10.1086/192291}
\end{barticle}
\endbibitem

\bibitem[\protect\citeauthoryear{{Marziani} et~al.}{2001}]{marzianietal01}
\begin{barticle}
\bauthor{\bsnm{{Marziani}}, \binits{P.}},
\bauthor{\bsnm{{Sulentic}}, \binits{J.W.}},
\bauthor{\bsnm{{Zwitter}}, \binits{T.}},
\bauthor{\bsnm{{Dultzin-Hacyan}}, \binits{D.}},
\bauthor{\bsnm{{Calvani}}, \binits{M.}}:
\bjtitle{ApJ}
\bvolume{558},
\bfpage{553}
(\byear{2001}).
\arxivurl{arXiv:astro-ph/0105343}.
doi:\doiurl{10.1086/322286}
\end{barticle}
\endbibitem

\bibitem[\protect\citeauthoryear{{Marziani} et~al.}{2003a}]{marzianietal03a}
\begin{barticle}
\bauthor{\bsnm{{Marziani}}, \binits{P.}},
\bauthor{\bsnm{{Sulentic}}, \binits{J.W.}},
\bauthor{\bsnm{{Zamanov}}, \binits{R.}},
\bauthor{\bsnm{{Calvani}}, \binits{M.}},
\bauthor{\bsnm{{Dultzin-Hacyan}}, \binits{D.}},
\bauthor{\bsnm{{Bachev}}, \binits{R.}},
\bauthor{\bsnm{{Zwitter}}, \binits{T.}}:
\bjtitle{ApJS}
\bvolume{145},
\bfpage{199}
(\byear{2003}a).
doi:\doiurl{10.1086/346025}
\end{barticle}
\endbibitem

\bibitem[\protect\citeauthoryear{{Marziani} et~al.}{2003b}]{marzianietal03b}
\begin{barticle}
\bauthor{\bsnm{{Marziani}}, \binits{P.}},
\bauthor{\bsnm{{Zamanov}}, \binits{R.K.}},
\bauthor{\bsnm{{Sulentic}}, \binits{J.W.}},
\bauthor{\bsnm{{Calvani}}, \binits{M.}}:
\bjtitle{MNRAS}
\bvolume{345},
\bfpage{1133}
(\byear{2003}b).
\arxivurl{arXiv:astro-ph/0307367}.
doi:\doiurl{10.1046/j.1365-2966.2003.07033.x}
\end{barticle}
\endbibitem

\bibitem[\protect\citeauthoryear{{Marziani} et~al.}{2009}]{marzianietal09}
\begin{barticle}
\bauthor{\bsnm{{Marziani}}, \binits{P.}},
\bauthor{\bsnm{{Sulentic}}, \binits{J.W.}},
\bauthor{\bsnm{{Stirpe}}, \binits{G.M.}},
\bauthor{\bsnm{{Zamfir}}, \binits{S.}},
\bauthor{\bsnm{{Calvani}}, \binits{M.}}:
\bjtitle{A\&Ap}
\bvolume{495},
\bfpage{83}
(\byear{2009}).
\arxivurl{0812.0251}.
doi:\doiurl{10.1051/\-0004-\-6361:200810764}
\end{barticle}
\endbibitem

\bibitem[\protect\citeauthoryear{{Marziani} et~al.}{2010}]{marzianietal10}
\begin{barticle}
\bauthor{\bsnm{{Marziani}}, \binits{P.}},
\bauthor{\bsnm{{Sulentic}}, \binits{J.W.}},
\bauthor{\bsnm{{Negrete}}, \binits{C.A.}},
\bauthor{\bsnm{{Dultzin}}, \binits{D.}},
\bauthor{\bsnm{{Zamfir}}, \binits{S.}},
\bauthor{\bsnm{{Bachev}}, \binits{R.}}:
\bjtitle{\mnras}
\bvolume{409},
\bfpage{1033}
(\byear{2010}).
\arxivurl{1007.3187}.
doi:\doiurl{10.1111/j.1365-2966.2010.17357.x}
\end{barticle}
\endbibitem

\bibitem[\protect\citeauthoryear{{Mar\-ziani} et~ al.}{2013a}]{marzianietal13a}
\begin{botherref}
\oauthor{\bsnm{{Marziani}}, \binits{P.}},
\oauthor{\bsnm{{Sulentic}}, \binits{J.W.}},
\oauthor{\bsnm{{Plauchu-Frayn}}, \binits{I.}},
\oauthor{\bsnm{{del Olmo}}, \binits{A.}}:
AAp
(2013a).
\arxivurl{1305.1096}
\end{botherref}
\endbibitem

\bibitem[\protect\citeauthoryear{{Marziani} et~al.}{2013b}]{marzianietal13}
\begin{botherref}
\oauthor{\bsnm{{Marziani}}, \binits{P.}},
\oauthor{\bsnm{{Sulentic}}, \binits{J.W.}},
\oauthor{\bsnm{{Plauchu-Frayn}}, \binits{I.}},
\oauthor{\bsnm{{del Olmo}}, \binits{A.}}:
ApJ
\textbf{764}(150)
(2013b).
\arxivurl{1301.0520}
\end{botherref}
\endbibitem

\bibitem[\protect\citeauthoryear{{Mathews}}{1993}]{mathews93}
\begin{barticle}
\bauthor{\bsnm{{Mathews}}, \binits{W.G.}}:
\bjtitle{ApJL}
\bvolume{412},
\bfpage{17}
(\byear{1993}).
doi:\doiurl{10.1086/186929}
\end{barticle}
\endbibitem

\bibitem[\protect\citeauthoryear{{McLure} and {Dunlop}}{2002}]{mcluredunlop02}
\begin{barticle}
\bauthor{\bsnm{{McLure}}, \binits{R.J.}},
\bauthor{\bsnm{{Dunlop}}, \binits{J.S.}}:
\bjtitle{\mnras}
\bvolume{331},
\bfpage{795}
(\byear{2002}).
\arxivurl{astro-ph/0108417}.
doi:\doiurl{10.1046/j.1365-8711.2002.05236.x}
\end{barticle}
\endbibitem

\bibitem[\protect\citeauthoryear{{Negrete} et~al.}{2012}]{negreteetal12}
\begin{barticle}
\bauthor{\bsnm{{Negrete}}, \binits{A.}},
\bauthor{\bsnm{{Dultzin}}, \binits{D.}},
\bauthor{\bsnm{{Marziani}}, \binits{P.}},
\bauthor{\bsnm{{Sulentic}}, \binits{J.}}:
\bjtitle{ApJ}
\bvolume{757},
\bfpage{62}
(\byear{2012}).
\arxivurl{1107.3188}
\end{barticle}
\endbibitem

\bibitem[\protect\citeauthoryear{{Netzer}}{1977}]{netzer77}
\begin{barticle}
\bauthor{\bsnm{{Netzer}}, \binits{H.}}:
\bjtitle{\mnras}
\bvolume{181},
\bfpage{89}
(\byear{1977})
\end{barticle}
\endbibitem

\bibitem[\protect\citeauthoryear{{Netzer} and
  {Marziani}}{2010}]{netzermarziani10}
\begin{barticle}
\bauthor{\bsnm{{Netzer}}, \binits{H.}},
\bauthor{\bsnm{{Marziani}}, \binits{P.}}:
\bjtitle{\apj}
\bvolume{724},
\bfpage{318}
(\byear{2010}).
\arxivurl{1006.3553}.
doi:\doiurl{10.1088/0004-637X/724/1/318}
\end{barticle}
\endbibitem

\bibitem[\protect\citeauthoryear{{Osterbrock}}{1978}]{osterbrock78}
\begin{barticle}
\bauthor{\bsnm{{Osterbrock}}, \binits{D.E.}}:
\bjtitle{\physscr}
\bvolume{17},
\bfpage{285}
(\byear{1978}).
doi:\doiurl{10.1088/\-0031\--8949/17/3/024}
\end{barticle}
\endbibitem

\bibitem[\protect\citeauthoryear{{Osterbrock}}{1979}]{osterbrock79}
\begin{barticle}
\bauthor{\bsnm{{Osterbrock}}, \binits{D.E.}}:
\bjtitle{\pasp}
\bvolume{91},
\bfpage{608}
(\byear{1979})
\end{barticle}
\endbibitem

\bibitem[\protect\citeauthoryear{{Osterbrock}}{1981}]{osterbrock81}
\begin{barticle}
\bauthor{\bsnm{{Osterbrock}}, \binits{D.E.}}:
\bjtitle{\apj}
\bvolume{249},
\bfpage{462}
(\byear{1981}).
doi:\doiurl{10.\-1086/\-159306}
\end{barticle}
\endbibitem

\bibitem[\protect\citeauthoryear{{Pappa} et~al.}{2001}]{pappaetal01}
\begin{barticle}
\bauthor{\bsnm{{Pappa}}, \binits{A.}},
\bauthor{\bsnm{{Georgantopoulos}}, \binits{I.}},
\bauthor{\bsnm{{Stewart}}, \binits{G.C.}},
\bauthor{\bsnm{{Zezas}}, \binits{A.L.}}:
\bjtitle{\mnras}
\bvolume{326},
\bfpage{995}
(\byear{2001}).
\arxivurl{astro-ph/0104061}.
doi:\doiurl{10.1046/j.1365-8711.2001.04609.x}
\end{barticle}
\endbibitem

\bibitem[\protect\citeauthoryear{{Peach}}{1981}]{peach81}
\begin{barticle}
\bauthor{\bsnm{{Peach}}, \binits{G.}}:
\bjtitle{Advances in Physics}
\bvolume{30},
\bfpage{367}
(\byear{1981}).
doi:\doiurl{10.\-1080/\-00018738100101467}
\end{barticle}
\endbibitem

\bibitem[\protect\citeauthoryear{{Peterson} and
  {Wandel}}{1999}]{petersonwandel99}
\begin{barticle}
\bauthor{\bsnm{{Peterson}}, \binits{B.M.}},
\bauthor{\bsnm{{Wandel}}, \binits{A.}}:
\bjtitle{\apjl}
\bvolume{521},
\bfpage{95}
(\byear{1999}).
\arxivurl{arXiv:astro-ph/9905382}.
doi:\doiurl{10.1086/312190}
\end{barticle}
\endbibitem

\bibitem[\protect\citeauthoryear{{Petrov}}{2009}]{petrov09}
\begin{barticle}
\bauthor{\bsnm{{Petrov}}, \binits{G.P.}}:
\bjtitle{Bulgarian Astronomical Journal}
\bvolume{11},
\bfpage{79}
(\byear{2009})
\end{barticle}
\endbibitem

\bibitem[\protect\citeauthoryear{{Richards} et~al.}{2011}]{richardsetal11}
\begin{barticle}
\bauthor{\bsnm{{Richards}}, \binits{G.T.}},
\bauthor{\bsnm{{Kruczek}}, \binits{N.E.}},
\bauthor{\bsnm{{Gallagher}}, \binits{S.C.}},
\bauthor{\bsnm{{Hall}}, \binits{P.B.}},
\bauthor{\bsnm{{Hewett}}, \binits{P.C.}},
\bauthor{\bsnm{{Leighly}}, \binits{K.M.}},
\bauthor{\bsnm{{Deo}}, \binits{R.P.}},
\bauthor{\bsnm{{Kratzer}}, \binits{R.M.}},
\bauthor{\bsnm{{Shen}}, \binits{Y.}}:
\bjtitle{\aj}
\bvolume{141},
\bfpage{167}
(\byear{2011}).
\arxivurl{1011.2282}.
doi:\doiurl{10.1088/0004-6256/141/5/167}
\end{barticle}
\endbibitem

\bibitem[\protect\citeauthoryear{{Robinson}}{1995}]{robinson95}
\begin{barticle}
\bauthor{\bsnm{{Robinson}}, \binits{A.}}:
\bjtitle{\mnras}
\bvolume{272},
\bfpage{647}
(\byear{1995})
\end{barticle}
\endbibitem

\bibitem[\protect\citeauthoryear{{Rokaki} et~al.}{2003}]{rokakietal03}
\begin{barticle}
\bauthor{\bsnm{{Rokaki}}, \binits{E.}},
\bauthor{\bsnm{{Lawrence}}, \binits{A.}},
\bauthor{\bsnm{{Economou}}, \binits{F.}},
\bauthor{\bsnm{{Mastichiadis}}, \binits{A.}}:
\bjtitle{\mnras}
\bvolume{340},
\bfpage{1298}
(\byear{2003}).
\arxivurl{arXiv:astro-ph/0301405}.
doi:\doiurl{10.1046/j.1365-8711.2003.06414.x}
\end{barticle}
\endbibitem

\bibitem[\protect\citeauthoryear{{Shadmehri}}{2015}]{shadmehri15}
\begin{barticle}
\bauthor{\bsnm{{Shadmehri}}, \binits{M.}}:
\bjtitle{\mnras}
\bvolume{451},
\bfpage{3671}
(\byear{2015}).
\arxivurl{1506.00247}.
doi:\doiurl{10.1093/mnras/stv1212}
\end{barticle}
\endbibitem

\bibitem[\protect\citeauthoryear{{Shields} and {McKee}}{1981}]{shieldsmckee81}
\begin{barticle}
\bauthor{\bsnm{{Shields}}, \binits{G.A.}},
\bauthor{\bsnm{{McKee}}, \binits{C.F.}}:
\bjtitle{\apjl}
\bvolume{246},
\bfpage{57}
(\byear{1981}).
doi:\doiurl{10.1086/183552}
\end{barticle}
\endbibitem

\bibitem[\protect\citeauthoryear{{Shuder}}{1981}]{shuder81}
\begin{barticle}
\bauthor{\bsnm{{Shuder}}, \binits{J.M.}}:
\bjtitle{\apj}
\bvolume{244},
\bfpage{12}
(\byear{1981}).
doi:\doiurl{10.1086/\-158678}
\end{barticle}
\endbibitem

\bibitem[\protect\citeauthoryear{{Sillanpaa} et~al.}{1988}]{sillanpaaetal88}
\begin{barticle}
\bauthor{\bsnm{{Sillanpaa}}, \binits{A.}},
\bauthor{\bsnm{{Haarala}}, \binits{S.}},
\bauthor{\bsnm{{Valtonen}}, \binits{M.J.}},
\bauthor{\bsnm{{Sundelius}}, \binits{B.}},
\bauthor{\bsnm{{Byrd}}, \binits{G.G.}}:
\bjtitle{\apj}
\bvolume{325},
\bfpage{628}
(\byear{1988}).
doi:\doiurl{10.1086/\-166033}
\end{barticle}
\endbibitem

\bibitem[\protect\citeauthoryear{{Smith}}{1980}]{smith80}
\begin{barticle}
\bauthor{\bsnm{{Smith}}, \binits{H.E.}}:
\bjtitle{\apjl}
\bvolume{241},
\bfpage{137}
(\byear{1980}).
doi:\doiurl{10.\-1086/\-183377}
\end{barticle}
\endbibitem

\bibitem[\protect\citeauthoryear{{Smith} et~al.}{2005}]{smithetal05}
\begin{barticle}
\bauthor{\bsnm{{Smith}}, \binits{J.E.}},
\bauthor{\bsnm{{Robinson}}, \binits{A.}},
\bauthor{\bsnm{{Young}}, \binits{S.}},
\bauthor{\bsnm{{Axon}}, \binits{D.J.}},
\bauthor{\bsnm{{Corbett}}, \binits{E.A.}}:
\bjtitle{\mnras}
\bvolume{359},
\bfpage{846}
(\byear{2005}).
\arxivurl{astro-ph/0501640}.
doi:\doiurl{10.1111/j.1365-2966.2005.08895.x}
\end{barticle}
\endbibitem

\bibitem[\protect\citeauthoryear{{Snedden} and
  {Gaskell}}{2007}]{sneddengaskell07}
\begin{barticle}
\bauthor{\bsnm{{Snedden}}, \binits{S.A.}},
\bauthor{\bsnm{{Gaskell}}, \binits{C.M.}}:
\bjtitle{ApJ}
\bvolume{669},
\bfpage{126}
(\byear{2007}).
doi:\doiurl{10.1086/521290}
\end{barticle}
\endbibitem

\bibitem[\protect\citeauthoryear{{Stirpe}}{1990}]{stirpe90}
\begin{barticle}
\bauthor{\bsnm{{Stirpe}}, \binits{G.M.}}:
\bjtitle{A\&ApS}
\bvolume{85},
\bfpage{1049}
(\byear{1990})
\end{barticle}
\endbibitem

\bibitem[\protect\citeauthoryear{{Strateva} et~al.}{2003}]{stratevaetal03}
\begin{barticle}
\bauthor{\bsnm{{Strateva}}, \binits{I.V.}},
\bauthor{\bsnm{{Strauss}}, \binits{M.A.}},
\bauthor{\bsnm{{Hao}}, \binits{L.}},
\bauthor{\bsnm{{Schlegel}}, \binits{D.J.}},
\bauthor{\bsnm{{Hall}}, \binits{P.B.}},
\bauthor{\bsnm{{Gunn}}, \binits{J.E.}},
\bauthor{\bsnm{{Li}}, \binits{L.}},
\bauthor{\bsnm{{Ivezi{\'c}}}, \binits{{\v Z}.}},
\bauthor{\bsnm{{Richards}}, \binits{G.T.}},
\bauthor{\bsnm{{Zakamska}}, \binits{N.L.}},
\bauthor{\bsnm{{Voges}}, \binits{W.}},
\bauthor{\bsnm{{Anderson}}, \binits{S.F.}},
\bauthor{\bsnm{{Lupton}}, \binits{R.H.}},
\bauthor{\bsnm{{Schneider}}, \binits{D.P.}},
\bauthor{\bsnm{{Brinkmann}}, \binits{J.}},
\bauthor{\bsnm{{Nichol}}, \binits{R.C.}}:
\bjtitle{AJ}
\bvolume{126},
\bfpage{1720}
(\byear{2003}).
\arxivurl{arXiv:astro-ph/0307357}.
doi:\doiurl{10.1086/378367}
\end{barticle}
\endbibitem

\bibitem[\protect\citeauthoryear{{Sulentic} et~al.}{2011}]{sulenticetal11}
\begin{barticle}
\bauthor{\bsnm{{Sulentic}}, \binits{J.}},
\bauthor{\bsnm{{Marziani}}, \binits{P.}},
\bauthor{\bsnm{{Zamfir}}, \binits{S.}}:
\bjtitle{Baltic Astronomy}
\bvolume{20},
\bfpage{427}
(\byear{2011})
\end{barticle}
\endbibitem

\bibitem[\protect\citeauthoryear{{Sulentic}}{1989}]{sulentic89}
\begin{barticle}
\bauthor{\bsnm{{Sulentic}}, \binits{J.W.}}:
\bjtitle{\apj}
\bvolume{343},
\bfpage{54}
(\byear{1989}).
doi:\doiurl{10.1086/\-167684}
\end{barticle}
\endbibitem

\bibitem[\protect\citeauthoryear{{Sulentic} et~al.}{2000}]{sulenticetal00a}
\begin{barticle}
\bauthor{\bsnm{{Sulentic}}, \binits{J.W.}},
\bauthor{\bsnm{{Marziani}}, \binits{P.}},
\bauthor{\bsnm{{Dultzin-Hacyan}}, \binits{D.}}:
\bjtitle{ARA\&A}
\bvolume{38},
\bfpage{521}
(\byear{2000}).
doi:\doiurl{10.1146/annurev.astro.38.1.521}
\end{barticle}
\endbibitem

\bibitem[\protect\citeauthoryear{{Sulentic} et~al.}{1995}]{sulenticetal95}
\begin{barticle}
\bauthor{\bsnm{{Sulentic}}, \binits{J.W.}},
\bauthor{\bsnm{{Marziani}}, \binits{P.}},
\bauthor{\bsnm{{Dultzin-Hacyan}}, \binits{D.}},
\bauthor{\bsnm{{Calvani}}, \binits{M.}},
\bauthor{\bsnm{{Moles}}, \binits{M.}}:
\bjtitle{ApJL}
\bvolume{445},
\bfpage{85}
(\byear{1995}).
doi:\doiurl{10.1086/187896}
\end{barticle}
\endbibitem

\bibitem[\protect\citeauthoryear{{Sulentic} et~al.}{2002}]{sulenticetal02}
\begin{barticle}
\bauthor{\bsnm{{Sulentic}}, \binits{J.W.}},
\bauthor{\bsnm{{Marziani}}, \binits{P.}},
\bauthor{\bsnm{{Zamanov}}, \binits{R.}},
\bauthor{\bsnm{{Bachev}}, \binits{R.}},
\bauthor{\bsnm{{Calvani}}, \binits{M.}},
\bauthor{\bsnm{{Dultzin-Hacyan}}, \binits{D.}}:
\bjtitle{ApJL}
\bvolume{566},
\bfpage{71}
(\byear{2002}).
\arxivurl{arXiv:astro-ph/0201362}.
doi:\doiurl{10.1086/339594}
\end{barticle}
\endbibitem

\bibitem[\protect\citeauthoryear{{Sulentic} et~al.}{2003}]{sulenticetal03}
\begin{barticle}
\bauthor{\bsnm{{Sulentic}}, \binits{J.W.}},
\bauthor{\bsnm{{Zamfir}}, \binits{S.}},
\bauthor{\bsnm{{Marziani}}, \binits{P.}},
\bauthor{\bsnm{{Bachev}}, \binits{R.}},
\bauthor{\bsnm{{Calvani}}, \binits{M.}},
\bauthor{\bsnm{{Dultzin-Hacyan}}, \binits{D.}}:
\bjtitle{ApJL}
\bvolume{597},
\bfpage{17}
(\byear{2003}).
\arxivurl{arXiv:astro-ph/0309469}.
doi:\doiurl{10.1086/379754}
\end{barticle}
\endbibitem

\bibitem[\protect\citeauthoryear{{Sulentic} et~al.}{2007}]{sulenticetal07}
\begin{barticle}
\bauthor{\bsnm{{Sulentic}}, \binits{J.W.}},
\bauthor{\bsnm{{Bachev}}, \binits{R.}},
\bauthor{\bsnm{{Marziani}}, \binits{P.}},
\bauthor{\bsnm{{Negrete}}, \binits{C.A.}},
\bauthor{\bsnm{{Dultzin}}, \binits{D.}}:
\bjtitle{ApJ}
\bvolume{666},
\bfpage{757}
(\byear{2007}).
\arxivurl{0705.1895}.
doi:\doiurl{10.10\-86\-/\-51\-99\-16}
\end{barticle}
\endbibitem

\bibitem[\protect\citeauthoryear{{Sulentic} et~al.}{2008}]{sulenticetal08}
\begin{bchapter}
\bauthor{\bsnm{{Sulentic}}, \binits{J.W.}},
\bauthor{\bsnm{{Zamfir}}, \binits{S.}},
\bauthor{\bsnm{{Marziani}}, \binits{P.}},
\bauthor{\bsnm{{Dultzin}}, \binits{D.}}:
In: \bbtitle{Revista Mexicana de Astronomia y Astrofisica Conference Series}.
\bsertitle{Revista Mexicana de Astronomia y Astrofisica Conference Series},
vol. \bseriesno{32},
p. \bfpage{51}
(\byear{2008})
\end{bchapter}
\endbibitem

\bibitem[\protect\citeauthoryear{{Sulentic} et~al.}{2012}]{sulenticetal12}
\begin{barticle}
\bauthor{\bsnm{{Sulentic}}, \binits{J.W.}},
\bauthor{\bsnm{{Marziani}}, \binits{P.}},
\bauthor{\bsnm{{Zamfir}}, \binits{S.}},
\bauthor{\bsnm{{Meadows}}, \binits{Z.A.}}:
\bjtitle{\apjl}
\bvolume{752},
\bfpage{7}
(\byear{2012}).
\arxivurl{1203.5992}.
doi:\doiurl{10.1088/\-2041-8205/752/1/L7}
\end{barticle}
\endbibitem

\bibitem[\protect\citeauthoryear{{Tytler} and {Fan}}{1992}]{tytlerfan92}
\begin{barticle}
\bauthor{\bsnm{{Tytler}}, \binits{D.}},
\bauthor{\bsnm{{Fan}}, \binits{X.-M.}}:
\bjtitle{ApJS}
\bvolume{79},
\bfpage{1}
(\byear{1992}).
doi:\doiurl{10.1086/19\-16\-42}
\end{barticle}
\endbibitem

\bibitem[\protect\citeauthoryear{{Volonteri} et~al.}{2009}]{volonterietal09}
\begin{barticle}
\bauthor{\bsnm{{Volonteri}}, \binits{M.}},
\bauthor{\bsnm{{Miller}}, \binits{J.M.}},
\bauthor{\bsnm{{Dotti}}, \binits{M.}}:
\bjtitle{\apjl}
\bvolume{703},
\bfpage{86}
(\byear{2009}).
\arxivurl{0903.3947}.
doi:\doiurl{10.1088/0004-637X/\-703/1/\-L86}
\end{barticle}
\endbibitem

\bibitem[\protect\citeauthoryear{{Wang} et~al.}{2014}]{wangetal14a}
\begin{barticle}
\bauthor{\bsnm{{Wang}}, \binits{J.-M.}},
\bauthor{\bsnm{{Qiu}}, \binits{J.}},
\bauthor{\bsnm{{Du}}, \binits{P.}},
\bauthor{\bsnm{{Ho}}, \binits{L.C.}}:
\bjtitle{\apj}
\bvolume{797},
\bfpage{65}
(\byear{2014}).
\arxivurl{1410.5285}.
doi:\doiurl{10.1088/0004-637X/797/1/65}
\end{barticle}
\endbibitem

\bibitem[\protect\citeauthoryear{{Wiese} and
  {Kelleher}}{1971}]{wiesekelleher71}
\begin{barticle}
\bauthor{\bsnm{{Wiese}}, \binits{W.L.}},
\bauthor{\bsnm{{Kelleher}}, \binits{D.E.}}:
\bjtitle{\apjl}
\bvolume{166},
\bfpage{59}
(\byear{1971}).
doi:\doiurl{10.1086/180739}
\end{barticle}
\endbibitem

\bibitem[\protect\citeauthoryear{{Wills} and {Browne}}{1986}]{willsbrowne86}
\begin{barticle}
\bauthor{\bsnm{{Wills}}, \binits{B.J.}},
\bauthor{\bsnm{{Browne}}, \binits{I.W.A.}}:
\bjtitle{\apj}
\bvolume{302},
\bfpage{56}
(\byear{1986}).
doi:\doiurl{10.1086/163973}
\end{barticle}
\endbibitem

\bibitem[\protect\citeauthoryear{{Wolter} et~al.}{2005}]{wolteretal05}
\begin{barticle}
\bauthor{\bsnm{{Wolter}}, \binits{A.}},
\bauthor{\bsnm{{Gioia}}, \binits{I.M.}},
\bauthor{\bsnm{{Henry}}, \binits{J.P.}},
\bauthor{\bsnm{{Mullis}}, \binits{C.R.}}:
\bjtitle{\aap}
\bvolume{444},
\bfpage{165}
(\byear{2005}).
\arxivurl{astro-ph/0510045}.
doi:\doiurl{10.1051/0004-6361:20053441}
\end{barticle}
\endbibitem

\bibitem[\protect\citeauthoryear{{You} et~al.}{1986}]{youetal86}
\begin{barticle}
\bauthor{\bsnm{{You}}, \binits{J.-H.}},
\bauthor{\bsnm{{Cheng}}, \binits{F.-H.}},
\bauthor{\bsnm{{Cheng}}, \binits{F.-Z.}},
\bauthor{\bsnm{{Kiang}}, \binits{T.}}:
\bjtitle{Phys. Rev. A}
\bvolume{34},
\bfpage{3015}
(\byear{1986}).
doi:\doiurl{10.1103/PhysRevA.34.3015}
\end{barticle}
\endbibitem

\bibitem[\protect\citeauthoryear{{Zamfir} et~al.}{2008}]{zamfiretal08}
\begin{barticle}
\bauthor{\bsnm{{Zamfir}}, \binits{S.}},
\bauthor{\bsnm{{Sulentic}}, \binits{J.W.}},
\bauthor{\bsnm{{Marziani}}, \binits{P.}}:
\bjtitle{MNRAS}
\bvolume{387},
\bfpage{856}
(\byear{2008}).
\arxivurl{0804.0788}.
doi:\doiurl{10.1111/j.1365-2966.2008.13290.x}
\end{barticle}
\endbibitem

\bibitem[\protect\citeauthoryear{{Zamfir} et~al.}{2010}]{zamfiretal10}
\begin{barticle}
\bauthor{\bsnm{{Zamfir}}, \binits{S.}},
\bauthor{\bsnm{{Sulentic}}, \binits{J.W.}},
\bauthor{\bsnm{{Marziani}}, \binits{P.}},
\bauthor{\bsnm{{Dultzin}}, \binits{D.}}:
\bjtitle{\mnras}
\bvolume{403},
\bfpage{1759}
(\byear{2010}).
\arxivurl{0912.4306}.
doi:\doiurl{10.1111/j.1365-2966.2009.16236.x}
\end{barticle}
\endbibitem

\bibitem[\protect\citeauthoryear{{Zhang}}{2011}]{zhang11}
\begin{barticle}
\bauthor{\bsnm{{Zhang}}, \binits{X.-G.}}:
\bjtitle{\mnras}
\bvolume{416},
\bfpage{2857}
(\byear{2011}).
\arxivurl{1107.0455}.
doi:\doiurl{10.1111/j.1365-2966.2011.19234.x}
\end{barticle}
\endbibitem

\end{thebibliography}

\end{document}